\documentclass[apj]{emulateapj}
\newcommand\msun{\ensuremath{M_\sun}}
\newcommand\teff{\ensuremath{T_{\rm eff}}}
\newcommand\logg{\ensuremath{\log g}}
\newcommand\tcool{\ensuremath{\tau_{\rm cool}}}
\newcommand\mf{\ensuremath{M_{\rm f}}}

\newcommand\mwd{\ensuremath{M_{\rm WD}}}

\newcommand\ubv{\ensuremath{U\!BV}}
\newcommand\ebv{\ensuremath{E(\bv)}}

\journalinfo{}
\slugcomment{Accepted for publication in the Astronomical Journal}
\shorttitle{White dwarfs in NGC 6633 \& NGC 7063}
\shortauthors{Williams \& Bolte}

\begin{document}

\title{A Photometric and Spectroscopic Search for White Dwarfs in the
Open Clusters NGC 6633 and NGC 7063\footnotemark[1]}
\footnotetext[1]{Some of the
data presented herein were obtained at the W.~M.~Keck Observatory,
which is operated as a scientific partnership among the California
Institute of Technology, the University of California and the National
Aeronautics and Space Administration. The Observatory was made
possible by the generous financial support of the W.~M.~Keck
Foundation.}

\author{Kurtis A. Williams}
\email{kurtis@astro.as.utexas.edu}
\affil{NSF Astronomy and Astrophysics Postdoctoral Fellow}
\affil{University of Texas, 1 University Station, C1400, Austin, TX 78712}
\and
\author{Michael Bolte}
\email{bolte@ucolick.org}
\affil{UCO/Lick Observatory, University of California, Santa Cruz, CA 95064}

\begin{abstract}
We present photometric and spectroscopic studies of the white dwarf
(WD) populations in the intermediate-age open clusters NGC 6633 and
NGC 7063 as part of the ongoing Lick-Arizona White Dwarf Survey
(LAWDS).  Using wide-field CCD imaging, we locate 41 candidate WDs in
the two cluster fields: 32 in NGC 6633, and 9 in NGC 7063.
Spectroscopic observations confirm 13 of these candidates to be
bona-fide WDs.  We describe in detail our Balmer line fitting
technique for deriving effective temperatures and surface gravities
from optical DA WD spectra and apply the technique to the 11 DA WDs in
the sample.  Of these, only two DA WDs are at the cluster distance
moduli, one in each cluster.  Two more DAs lie 0.75 mag foreground to
NGC 6633, raising the possibility that these are double degenerate
systems in the cluster.  If nearly equal-mass binaries, both of these
systems likely have combined masses above the Chandrasekhar limit.  One DB WD
is found to be consistent with membership in NGC 6633, which would
make this the third confirmed He-atmosphere WD in an open cluster,
though further data are needed to confirm cluster
membership.  The WD consistent with membership in the cluster NGC 7063
has a low mass ($\approx 0.4\msun$), suggesting it may be a He-core WD
resulting from close binary evolution.  Three of the eleven
hydrogen-atmosphere WDs in this study are observed to have \ion{Ca}{2}
absorption; the number of DAZs in this study is consistent with
previous observations that $\sim 25\%$ of field WDs are DAZs.

\end{abstract}


\section{Introduction}
\setcounter{footnote}{1}
White dwarfs (WDs) represent the endpoint of stellar evolution for the
majority of stars.  As such, WDs present an opportunity to study
stellar populations long after stellar evolution is complete.  WDs in
star clusters are especially useful due to their residence in a
co-eval, single-metallicity population, allowing determination of
their progenitor masses.

Studies of WDs in open star clusters are often used to address the
relation between a star's main-sequence mass and that of its WD
remnant (the initial-final mass relation, or IFMR).  The IFMR
represents the integrated mass lost by a star over its entire
evolution.  An understanding of this relation is necessary for
understanding chemical enrichment and star formation efficiency in
galaxies \citep{2005MNRAS.361.1131F} and for understanding the origin
and evolution of hot gas in giant elliptical galaxies
\citep[e.g.,][]{1990ApJ...354..468M}.  The IFMR also plays a vital
role in the study of the field WD luminosity function (WDLF), as the
WDLF represents a convolution of the IFMR with the star formation
history of a stellar population.  The WDLF is one of the primary means
to determine the age of the Galactic thin disk
\citep[e.g.,][]{1987ApJ...315L..77W,1992ApJ...386..539W,1996Natur.382..692O,1998ApJ...497..294L,2005ASPC..334..131K,2005ASPC..334....3V},
and interest is building in attempts to determine the age of the thick
disk and the halo via similar methods
\citep[e.g.,][]{2005ASPC..334....3V}.

The first observational work on WDs in open star clusters was
presented by \citet{1980ApJ...235..992R}, followed closely by
\citet{1981PhDT.........1A,1982ApJ...255..245A}.  Both used
photographic plates to search for overdensities of faint blue objects
in intermediate-age open clusters (ages $\sim 100$ Myr to $\sim 1$
Gyr).  This work was followed by a series of papers by Koester \&
Reimers, who presented follow-up spectroscopy of many of these and
other photographic WD candidates
\citep{1981A&A....99L...8K,1985A&A...153..260K,1993A&A...275..479K,1996A&A...313..810K,1982A&A...116..341R,1988A&A...202...77R,1989A&A...218..118R,1994A&A...285..451R}.
The addition of spectroscopy is crucial. First, it allows the
unambiguous identification of the bona-fide WDs in the samples. In the
photometric surveys, candidate WDs have been identified
spectroscopically as hot horizontal-branch, main-sequence or subdwarf
stars, and QSOs.

Second, high quality spectra of the WDs allow the physical properties
of the WDs to be determined.  By fitting synthetic spectra to the
observed spectra, the effective temperature (\teff) and surface
gravity (\logg) of each WD is determined.  Using WD evolutionary
models, these quantities can be converted into a WD mass (\mwd) and
cooling time (\tcool).  Subtraction of \tcool ~from a cluster's age
results in the total age of the progenitor star.  From stellar
evolutionary models, the progenitor mass is thereby determined.
\citet{2000A&A...363..647W} uses data achieved via this method along
with theoretical stellar evolutionary models to derive a
semi-empirical IFMR.

\begin{deluxetable*}{lcccl}
\tablewidth{0pt}
\tablecolumns{5}
\tablecaption{Log of observations.\label{tab.obslog}}
\tablehead{\colhead{UT Date} & \colhead{Facility} &
  \colhead{Instrument} & \colhead{Seeing} & \colhead{Comments} \\
 & & & \colhead{(\arcsec)} & }
\startdata
1999 October 16-17 & CFHT & CFH12K & 0.7 & Archived images; see \citet{2001AJ....122..257K}\\
2001 August 22-23 & Keck I & LRIS-B & $0.9-1.0$ & Engineering-grade CCDs \\
2001 September 20 & Lick 3m & PFCam & 1.6 & Non-linear CCD response \\
2002 July 6-8 & Lick 1m & Nickel CCD & $2-5$ & photometric \\
2002 August 6-7 & Keck I & LRIS-B & $0.6 - 0.7$ & variable cirrus \\
2002 September 7 & Lick 3m & PFCam & 2.2 & photometric \\
2002 December 8-9 & Keck I & LRIS-B & $0.9-1.2$ & clear \\
2004 February 12 & Keck I & LRIS-B & 0.9 & clear \\
2005 November 26 & Keck I & LRIS-B & 0.9 & clear \\
\enddata
\end{deluxetable*}

The advent of wide-field CCD cameras and blue-sensitive spectrometers
on 8m -- 10m class telescopes now permits the study of cluster WDs
with higher accuracy and to larger distances.  The first such study is
that of \citet{2001ApJ...563..987C}, who use proper motions and
digital photometry to locate WDs in the core of Praesepe, which are
then followed up with spectroscopy at large-aperture telescopes.

We have undertaken a similar approach to detecting and analyzing WDs
in intermediate-age open clusters, with initial results presented in
\citet{2002PhDT........17W}.  The first results of our study, now
tagged with the somewhat clumsy and outdated (but IAU-approved)
appellation ``Lick-Arizona White Dwarf Survey (LAWDS),'' were seven
massive WDs in the open cluster \object{NGC 2168} (M35)
\citep{2004ApJ...615L..49W}.

At least two other groups are conducting concurrent open cluster WD
studies.  \citet{2005ApJ...618L.123K} present analysis of WDs in the
open cluster \object{NGC 2099} (M37) and are continuing this work in
other older open clusters (J.~Kalirai 2006, personal communication).
In addition, \citet{2004MNRAS.355L..39D,2006MNRAS.369..383D} have
analyzed several WDs in Praesepe not included in the
\citet{2001ApJ...563..987C} study.

In addition to leading to a more defined IFMR, the recent cluster
observations have revealed the somewhat surprising possibility that
the ratio of hydrogen-atmosphere WDs (spectral class DA) to
helium-atmosphere WDs (the non-DA spectral classes) may differ between
the field and cluster WD populations.  This effect was noted by
\citet{1994A&A...285..451R}, and was more recently explored by
\citet{2005ApJ...618L.129K}.  We note in \citet{2006ApJ...643L.127W}
that some of the apparent discrepancy may be due to the known
temperature dependence of the DA:non-DA ratio, though there is still a
possible deficiency of non-DA WDs in clusters as compared to the
field.  It is therefore important to analyze carefully any non-DA
spectral type WDs in open cluster fields.

In this paper, we present photometry and spectroscopy of WDs in two
intermediate-age, sparse open clusters, \object{NGC 6633} and
\object{NGC 7063}.  This paper also details our spectroscopic analysis
technique, which was only briefly mentioned in
\citet{2004ApJ...615L..49W}.  In \S 2, we describe the two
open clusters and our photometric observations thereof.
\S 3 describes our spectroscopic observations, while
\S 4 presents our spectral fitting routine and testing
of the routine in gory detail.  We present our results in
\S 5 and discuss these results in respect to the IFMR and
other open cluster WD issues.  In this section we also discuss several
interesting individual objects.

In this paper, we assume the solar metallicity value of
\citet[][$Z=0.013$]{2004A&A...417..751A} and the extinction curve of
\citet{1985ApJ...288..618R} with $R_V=3.1$.  We note that initial
analysis of these data were included in \citet{2002PhDT........17W};
all analysis presented in this paper supersedes that in the previous
work.

\section{Photometric Observations and Analysis\label{sec.phot}}

\bv ~images of NGC 6633 were obtained from the public CFHT archive of
the Canadian Astronomy Data Centre.  These images were taken as part
of the CFHT Open Star Cluster Survey \citep{2001AJ....122..257K} using
the CFH12K mosaic camera on UT 1999 October 16.  We reduced the data
using the IRAF\footnote{IRAF is distributed by the National Optical
Astronomy Observatories, which are operated by the Association of
Universities for Research in Astronomy, Inc., under cooperative
agreement with the National Science Foundation.} external package
CFH12K, an extension to the MSCRED mosaic reduction package.  The
twilight flat fields obtained on the night of observation were not
useful due to high stellar densities, so twilight flat field images
from subsequent nights were retrieved from the archive and used to
construct flat fields.  The data were bias-subtracted, trimmed,
flat-fielded, and resampled to a common pixel scale and coordinate
grid, the basic algorithm suggested by
\citet[][]{2002adaa.conf..309V}.  Due to the large number of bad
columns in CCD05 of CFH12K, data on this chip were ignored.

\ubv ~images of NGC 7063 were obtained using the Prime Focus Camera
(PFCam) on the Lick 3m telescope on 2001 September 21 and 2002
September 7.  PFCam uses a SITe $2048\times 2048$ backside-illuminated
CCD mounted above a prime focus field corrector and an atmospheric
dispersion corrector.  The unbinned pixel scale is 0\farcs296
pixel$^{-1}$ for a field of view $\approx 10\arcmin\times 10\arcmin$.
Our observations were made with $2\times 2$ binning.

For the 2001 observations, seeing was $\approx 1\farcs5$, though these
images suffered from a non-linear response of undetermined origin.
The resulting photometry was of sufficient quality to select objects
with UV-excess, but not for photometric measurements.  For the 2002
observations, seeing was $\approx 2\farcs2$ and the device response
was linear, so these observations were used for photometric
measurements.

The log of all observations is given in Table \ref{tab.obslog}.

\subsection{Photometric Measurements and Calibration}
Photometry was obtained via point-spread function (PSF) fitting using
DAOPHOT II \citep{1987PASP...99..191S}.  The PSF was allowed to vary
quadratically in both x- and y-coordinates.  As the stellar background
was crowded, stellar PSFs were subtracted from the image, and a second
detection iteration was performed.

Because of shutter timing errors with PFCam and uncertainty about the
weather during the CFHT imaging, \ubv ~imaging of portions of both
clusters was obtained with the Nickel 1-m telescope at Lick
Observatory on UT 2002 July 6-8. The Nickel CCD Camera has a SITe
$1024\times1024$-pixel thinned CCD; pixels were binned $2\times 2$ for
an effective scale of $0\farcs56$ pix$^{-1}$ and a $\approx
5\arcmin\times5 \arcmin$ field-of-view.  Images were trimmed,
bias-subtracted, and flat-fielded.  We used aperture photometry and
the DAOGROW routine to determine total magnitudes for each star.

These magnitudes were calibrated using aperture photometry of standard
stars from \citet{1992AJ....104..340L} fields taken at a variety of
times and airmasses.  These data were used to solve the following
transformation equations:
\begin{eqnarray}
\scriptstyle u  & \scriptstyle = & \scriptstyle U + 2.5\log t_{\rm exp} + A_0 + A_1 (U-B) + A_2(X-1.25) + A_3 T \label{eqn.u} \\
\scriptstyle b  & \scriptstyle = &\scriptstyle  B + 2.5\log t_{\rm exp} + B_0 + B_1 (B-V) + B_2(X-1.25) \label{eqn.b} \\
\scriptstyle v  & \scriptstyle = & \scriptstyle V + 2.5\log t_{\rm exp} + C_0 + C_1 (B-V) + C_2(X-1.25) \label{eqn.v} \, ,
\end{eqnarray}
where $u,\, b,\, v$ are the total instrumental magnitudes (including a
25 mag zeropoint offset included in the DAOphot routines), $U,\, B,\,
V$ are the standard magnitudes, $t_{\rm exp}$ is the exposure time,
$X$ is the airmass, and $T$ is the UT time of the observation, in
hours.  As initial $T$ terms for $B$ and $V$ transformations were
consistent with zero, these are not included in the final calibration.
Due to a lack of well-measured standards in $U$, only $U$ observations
taken on 2002 July 8 could be accurately calibrated.  The photometric
coefficients are given in Table \ref{tab.photcoeff}.

\begin{deluxetable*}{lccccc}
\tablecolumns{6}
\tablewidth{0pt}
\tablecaption{Photometric Transformation Equation Coefficients}
\tablehead{\colhead{UT Date} & \colhead{Filter} & \colhead{Zero Point} & 
           \colhead{Color Term} & \colhead{Airmass Term} & \colhead{Obs.~Time Term}
           \label{tab.photcoeff}}
\startdata
2002 Jul 6 & $B$ & $2.248\pm 0.003$ & $-0.083\pm 0.002$ & $0.223\pm 0.008$ & \nodata \\
           & $V$ & $1.986\pm 0.005$ & $ \phantom{-}0.060\pm 0.001$ & $0.136\pm 0.005$ & \nodata \\
2002 Jul 7 & $B$ & $2.611\pm 0.016$ & $-0.083\pm 0.002$ & $0.156\pm 0.038$ & \nodata \\
           & $V$ & $2.351\pm 0.009$ & $ \phantom{-}0.060\pm 0.001$ & $0.070\pm 0.064$ & \nodata \\
2002 Jul 8 & $U$ & $4.199\pm 0.009$ & $-0.061\pm 0.002$ & $0.511\pm 0.012$ & $0.006\pm 0.001$ \\
           & $B$ & $2.573\pm 0.002$ & $-0.083\pm 0.002$ & $0.248\pm 0.008$ & \nodata \\
           & $V$ & $2.330\pm 0.002$ & $ \phantom{-}0.060\pm 0.001$ & $0.145\pm 0.008$ & \nodata \\
\enddata
\end{deluxetable*}

These transformation equations were then applied to the star cluster
images, and stars with calibrated photometry were then used as local
standards in the CFHT and PFCam imaging to determine the zero points
and color terms.  For the CFHT, we calculated color terms of
$B_1=0.057\pm 0.029$ and $C_1 = -0.024\pm 0.029$.  For PFCam, the
corresponding color terms were $A_1=0.095\pm 0.009$, $B_1=-0.093\pm
0.005$, and $C_1=0.079\pm 0.005$.

\subsection{White Dwarf Candidate Selection}
\subsection{NGC 6633}

\object{NGC 6633} is a loose open cluster with recent age estimates
ranging from 426 Myr \citep{2002A&A...389..871D} to 630 Myr
\citep{Lynga87}, similar in age to or slightly younger than the
\object{Hyades} \citep[$\approx 625$ Myr;][]{1998A&A...331...81P}. The
cluster is slightly metal-poor ($[{\rm Fe}/{\rm H}]\approx -0.1$), has
a distance modulus of $(m-M)_0=8.01\pm 0.09$ (for a Pleiades distance
modulus of 5.6), and is significantly reddened, with $\ebv=0.165\pm
0.011$ \citep{2002MNRAS.336.1109J}.

\begin{figure}
\plotone{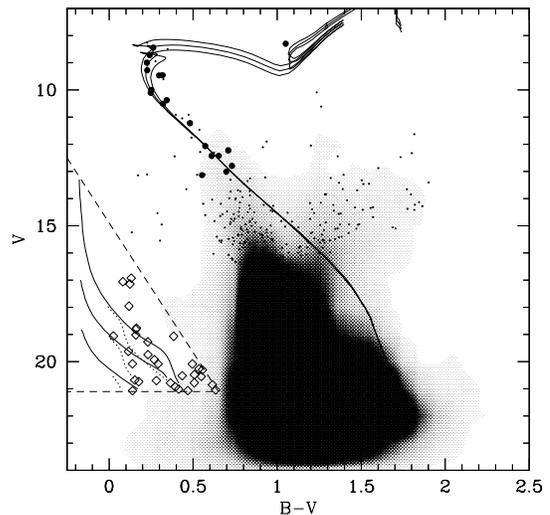}
\caption{Color-magnitude diagram for NGC 6633.  Grayscale reflects the
  relative stellar density.  Large filled circles are stars in the
  Nickel fields with proper motion membership probabilities $\geq
  50\%$ in \citet{2001A&A...376..441D}, while small dots are stars in
  the Nickel fields with lower membership probabilities or no
  measurement.  These points are also included in the grayscale.
  Solid isochrones are Padova isochrones for $\log t = 8.7,8.75,$ and
  8.8, interpolated to $Z=0.010$ and shifted to the cluster distance
  and reddening.  The dashed lines indicate candidate WD selection
  criteria, with the selected objects indicated as open diamonds.  WD
  cooling curves for cluster DAs (solid) and DBs (dashed) are given
  for $\log \tcool\leq 8.8$ and masses of 0.4\msun (top), 0.8\msun
  (middle), and 1.2\msun (bottom). No obvious WD cooling sequence is
  observed.
  \label{fig.n6633_cmd}}
\end{figure}

The color-magnitude diagram for NGC 6633 is shown in
Fig.~\ref{fig.n6633_cmd}. The isochrones shown are interpolated from
the $Z=0.008$ and $Z=0.019$ isochrones of \citet{2002A&A...391..195G},
assuming scaled solar abundance ratios and moderate convective
overshoot.  Interpolation was performed using a portion of the
StarFISH star-formation history code
\citep{2001ApJS..136...25H,2004AJ....127.1531H}, kindly provided by
J.~Harris.  Based on the apparent main-sequence turnoff in our new
photometry, we estimate a cluster age of $560^{+70}_{-60}$ Myr ($\log
t=8.75\pm 0.05$), in agreement with the published values.  A clear
main sequence can be seen down to $V\approx 15\,\,(M_V\approx 7)$. At
fainter magnitudes it becomes indistinguishable from the very large
field population.

Figure \ref{fig.n6633_cmd} also shows cooling curves for a range of WD
masses.  As no WD cooling sequence is obvious from photometry alone,
we use a broad color and photometry selection to identify WD
candidates.  The selection region includes DA and DB models at the
cluster distance and reddening for $\mwd\geq 0.4\msun$ \& $\tcool\leq
630$ Myr.  As we expect any cluster WDs to be younger than the cluster
(and therefore brighter than the faint $V$ limit), these photometric
selection criteria should include all potential cluster WDs.
Photometry of all 32 photometric WD candidates is given in Table
\ref{tab.cands.n6633}.

\tabletypesize{\scriptsize}
\begin{deluxetable*}{lccccccccl}
\tablecolumns{10}
\tablewidth{0pt}
\tablecaption{Candidate White Dwarfs in NGC 6633\label{tab.cands.n6633}}
\tablehead{\colhead{Object} & \colhead{RA} & \colhead{Dec} & \colhead{$V$} &
     \colhead{$\sigma_{\rm V}$} & \colhead{$B-V$} & \colhead{$\sigma_{B-V}$} & 
     \colhead{Obs. Date} & \colhead{ID} & \colhead{Comments}}
\startdata
\object{NGC 6633:LAWDS  1} & 18:28:16.7 & 6:34:10.8 & 17.06 & 0.02 & 0.08 & 0.03 & 2002 Aug 6  & B & Slitmask\\ 
\object{NGC 6633:LAWDS  2} & 18:27:13.6 & 6:19:54.5 & 17.14 & 0.02 & 0.12 & 0.03 & 2002 Aug 6  & A & RA 6633-7; Slitmask\\  
\object{NGC 6633:LAWDS  3} & 18:27:03.0 & 6:24:28.3 & 17.96 & 0.02 & 0.12 & 0.03 & 2002 Aug 6  & B & RA 6633-4; Slitmask\\ 
\object{NGC 6633:LAWDS  4} & 18:27:10.4 & 6:26:15.7 & 18.82 & 0.02 & 0.16 & 0.03 & 2001 Aug 22,23 & DA\tablenotemark{a} & RA 6633-5\\ 
\object{NGC 6633:LAWDS  5} & 18:26:14.9 & 6:29:01.5 & 19.06 & 0.02 & 0.03 & 0.03 & 2002 Aug 7  &  O & RA 6633-1; Slitmask\\
\object{NGC 6633:LAWDS  6} & 18:27:43.7 & 6:32:57.1 & 19.07 & 0.02 & 0.38 & 0.03 & 2002 Aug 7  &  A & Slitmask\\
\object{NGC 6633:LAWDS  7} & 18:27:49.9 & 6:20:51.8 & 19.27 & 0.02 & 0.23 & 0.03 & 2001 Aug 23 & DA\tablenotemark{a} & \\
\object{NGC 6633:LAWDS  8} & 18:27:23.4 & 6:19:49.8 & 19.74 & 0.02 & 0.23 & 0.03 & 2002 Aug 6  & DAZ & Slitmask\\
\object{NGC 6633:LAWDS  9} & 18:26:57.5 & 6:43:32.0 & 20.68 & 0.02 & 0.15 & 0.03 & 2001 Aug 23 &  A & \\
\object{NGC 6633:LAWDS 10} & 18:26:52.4 & 6:27:29.3 & 20.70 & 0.02 & 0.28 & 0.03 & 2002 Aug 6  & B/A& \\
\object{NGC 6633:LAWDS 11} & 18:26:29.1 & 6:43:22.3 & 20.74 & 0.02 & 0.18 & 0.03 & 2002 Aug 6  &  B & \\ 
\object{NGC 6633:LAWDS 12} & 18:28:49.9 & 6:26:12.0 & 20.79 & 0.02 & 0.37 & 0.03 & 2002 Aug 6  & DA & \\
\object{NGC 6633:LAWDS 13} & 18:27:14.9 & 6:20:04.1 & 21.07 & 0.02 & 0.14 & 0.03 & 2001 Aug 23 & DA?& Slitmask \\
\object{NGC 6633:LAWDS 14} & 18:27:12.2 & 6:21:35.8 & 16.93 & 0.02 & 0.13 & 0.03 & 2001 Aug 22 & DB & RA 6633-6; Slitmask\\
\object{NGC 6633:LAWDS 15} & 18:26:08.1 & 6:24:51.0 & 19.63 & 0.02 & 0.12 & 0.03 & 2001 Aug 23 & DAZ & \\
                  &            &           &       &      &       &       & 2002 Aug  7 &    & Slitmask\\
\object{NGC 6633:LAWDS 16} & 18:28:04.7 & 6:45:06.3 & 20.08 & 0.02 & 0.14 & 0.03 & 2002 Aug  7 & DB & \\
\object{NGC 6633:LAWDS 17} & 18:28:03.0 & 6:41:27.3 & 20.27 & 0.12 & 0.54 & 0.14 & \nodata & \nodata & \\
\object{NGC 6633:LAWDS 18} & 18:28:03.1 & 6:23:19.4 & 20.48 & 0.02 & 0.51 & 0.03 & \nodata & \nodata & \\
\object{NGC 6633:LAWDS 19} & 18:27:35.5 & 6:38:11.5 & 20.55 & 0.02 & 0.55 & 0.03 & 2002 Aug  7 & QSO& $z=1.968$; Slitmask \\
\object{NGC 6633:LAWDS 20} & 18:26:56.4 & 6:39:05.3 & 20.86 & 0.23 & 0.62 & 0.23 & \nodata & \nodata & \\
\object{NGC 6633:LAWDS 21} & 18:27:49.0 & 6:45:27.8 & 21.02 & 0.03 & 0.41 & 0.04 & \nodata & \nodata & \\
\object{NGC 6633:LAWDS 22} & 18:27:07.8 & 6:42:29.7 & 21.04 & 0.02 & 0.63 & 0.04 & \nodata & \nodata & \\
\object{NGC 6633:LAWDS 23} & 18:26:17.5 & 6:22:32.5 & 18.77 & 0.02 & 0.16 & 0.03 & 2002 Aug  7 &  O & RA 6633-2; Slitmask \\
\object{NGC 6633:LAWDS 24} & 18:27:36.7 & 6:26:42.7 & 19.02 & 0.04 & 0.16 & 0.06 & \nodata & \nodata & \\
\object{NGC 6633:LAWDS 25} & 18:26:13.6 & 6:31:05.5 & 19.93 & 0.02 & 0.27 & 0.03 & 2003 Apr  6 &  A & \\
\object{NGC 6633:LAWDS 26} & 18:27:17.4 & 6:22:55.9 & 20.09 & 0.02 & 0.50 & 0.03 & \nodata & \nodata & \\
\object{NGC 6633:LAWDS 27} & 18:27:12.3 & 6:21:02.1 & 20.08 & 0.02 & 0.29 & 0.03 & 2002 Aug  6 & DAZ & Slitmask\\
                  &            &           &       &      &      &      & 2003 Apr  6 & \nodata & \\
\object{NGC 6633:LAWDS 28} & 18:27:22.2 & 6:36:26.0 & 20.31 & 0.02 & 0.56 & 0.03 & \nodata & \nodata & \\
\object{NGC 6633:LAWDS 29} & 18:27:52.6 & 6:32:55.0 & 20.52 & 0.02 & 0.44 & 0.03 & 2002 Aug  7 & QSO& $z=1.643$; Slitmask \\
\object{NGC 6633:LAWDS 30} & 18:27:30.3 & 6:24:59.7 & 20.78 & 0.02 & 0.51 & 0.03 & \nodata & \nodata & \\
\object{NGC 6633:LAWDS 31} & 18:26:21.6 & 6:45:48.0 & 20.93 & 0.12 & 0.39 & 0.13 & \nodata & \nodata & \\ 
\object{NGC 6633:LAWDS 32} & 18:26:16.0 & 6:22:58.6 & 21.07 & 0.03 & 0.47 & 0.04 & 2002 Aug  7 &  A & Slitmask\\ 
\enddata
\tablecomments{Units of right ascension are hours, minutes and seconds, and 
units of declination are degrees, arcminutes, and arcseconds.  Coordinates 
are for Equinox J2000.0.  Identifications of B indicate non-WD
spectra with \ion{He}{1} and H lines; identifications of A indicate
non-WD spectra with H lines.}
\tablenotetext{a}{Potential double degenerate star}
\end{deluxetable*}

\subsection{NGC 7063}

\object{NGC 7063} is a poor open cluster with an age around 95-125 Myr
\citep{2002A&A...389..871D,2005A&A...438.1163K}, making it similar in
age to the \object{Pleiades}.  The intrinsic distance modulus is 9.19
with a foreground reddening $\ebv=0.09$ \citep{2002A&A...389..871D}.
We have found no metallicity measurements for this cluster, and so
make the assumption that it is solar.

\begin{figure*}
\plottwo{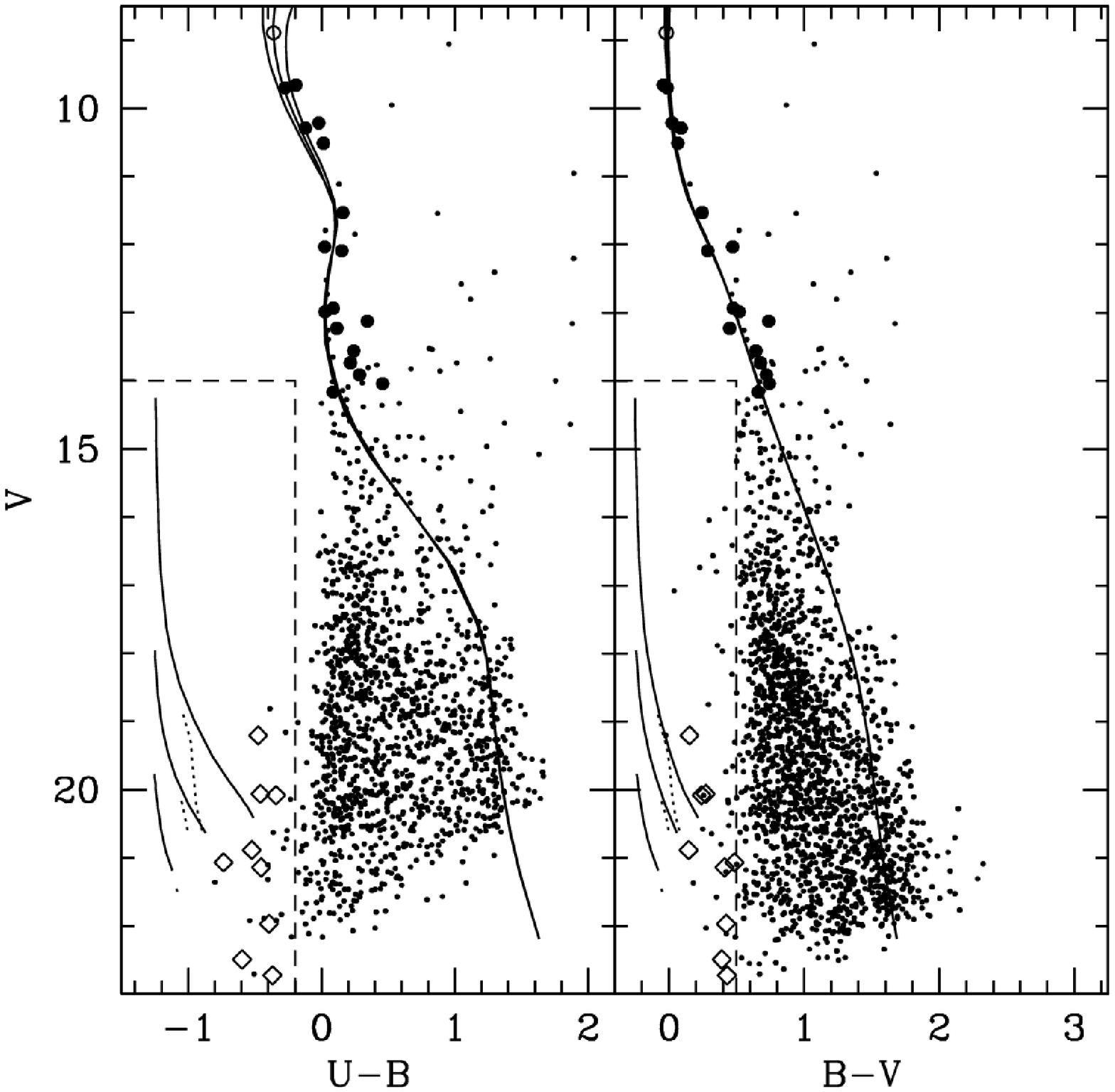}{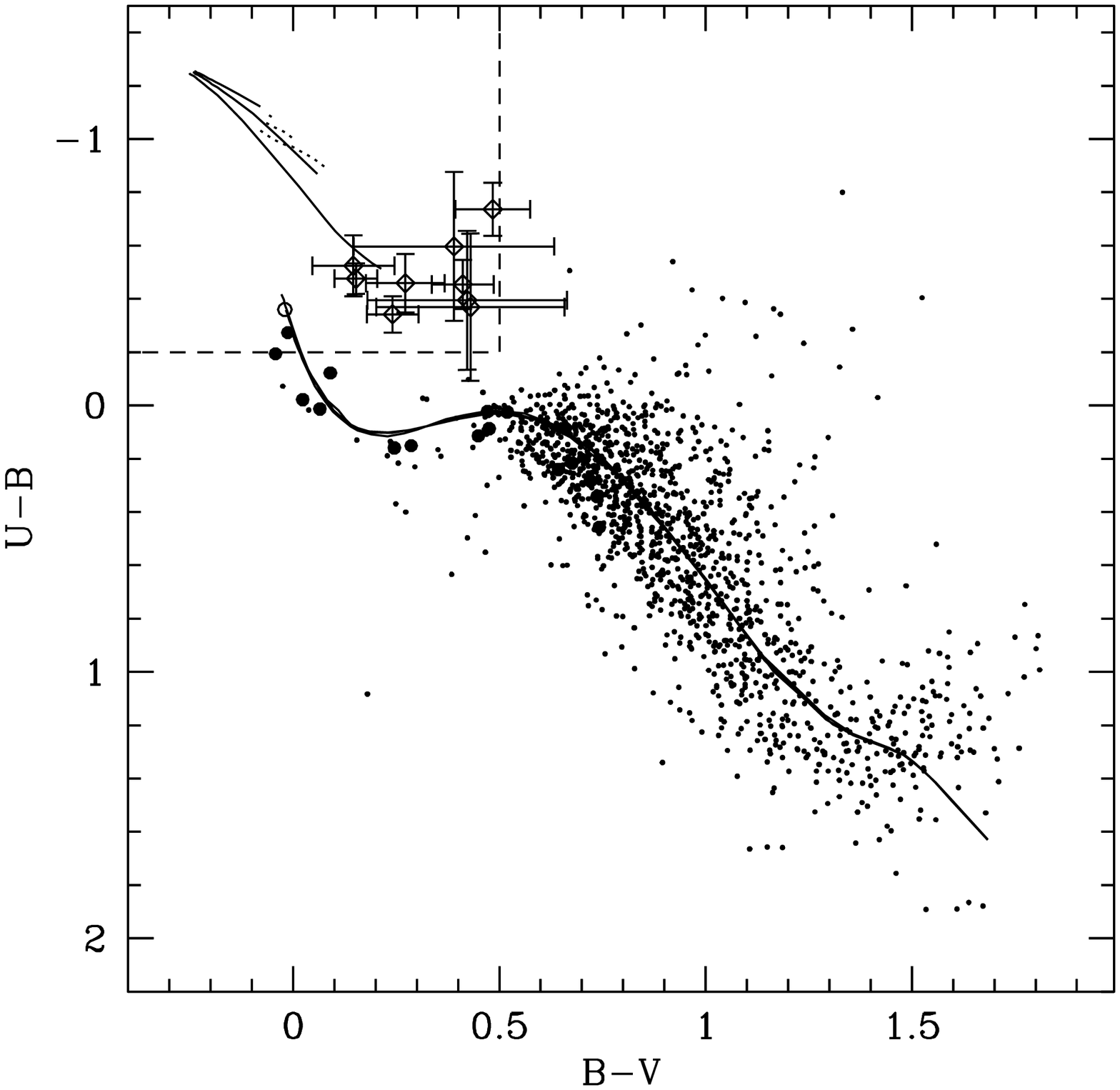}
\caption{Color-magnitude diagrams (left) and color-color plot (right)
for NGC 7063. Large filled points indicate stars with proper motion
membership probabilities $\geq 50\%$ in \citet{2001A&A...376..441D}.
Solid lines are$Z=0.013$ Padova isochrones of ages $\log t = 8.0,8.1,$
and 8.2. Only the $\log t=8.1$ isochrone is plotted in the color-color
diagram.  The open circle is HD 203921 from
\citet{1961PUSNO..17..343H}, saturated on our images.  The dashed
lines indicate candidate WD selection criteria, with the selected
objects indicated as open diamonds.  WD cooling curves for cluster DAs
(solid) and DBs (dashed) are given for $\log \tcool\leq 8.2$ and
masses of 0.4\msun (top), 0.8\msun (middle), and 1.2\msun (bottom). No
obvious WD cooling sequence is observed. \label{fig.n7063_cmd}}
\end{figure*}

The $\ub,\,V$ and $\bv,\, V$ CMDs for NGC 7063 are shown in Figure
\ref{fig.n7063_cmd}, along with $Z=0.013$ isochrones at the cluster
distance and reddening.  The isochrones are seen to give a good fit to
proper motion cluster members, and the best cluster age, based mainly
on previously-published photometry of \object{HD 203921}
\citep{1961PUSNO..17..343H}, is 125 Myr ($\log \tau = 8.1\pm 0.1$).

Candidate white dwarfs were selected using both \ub ~and \bv
~criteria, as shown in the color-color diagram of NGC 7063 (Figure
\ref{fig.n7063_cmd}).  Comparison of the color-color diagram with the
color-magnitude diagrams shows the utility of multiple colors in
selecting WD candidates, as many objects that would be selected as WD
candidates given a single color are eliminated based on two-color
photometry.  In total, nine WD candidates are identified, none of
which have photometry consistent with a high-mass WD at the cluster
distance. Photometry of all nine candidates is given in Table
\ref{tab.cands.n7063}.

\begin{deluxetable*}{lccccccccccl}
\tablecolumns{12}
\tablewidth{0pt}
\tablecaption{Candidate White Dwarfs in NGC 7063\label{tab.cands.n7063}}
\tablehead{\colhead{Object} & \colhead{RA} & \colhead{Dec} & \colhead{$V$} &
     \colhead{$\sigma_{V}$} & \colhead{$B-V$} & \colhead{$\sigma_{B-V}$} & 
     \colhead{$U-B$} & \colhead{$\sigma_{U-B}$} & 
     \colhead{Obs. Date} & \colhead{ID} & \colhead{Comments}}
\startdata
\object{NGC 7063:LAWDS  1} & 21:24:10.3 & +36:26:02.2 & 20.03 & 0.04 & 0.24 & 0.06 & -0.34 & 0.07 & 2002 Aug  6 & DA & Slitmask \\
                  &            &             &       &      &      &      &       &      & 2002 Dec  8 &    & \\
\object{NGC 7063:LAWDS  2} & 21:24:21.5 & +36:26:00.7 & 19.20 & 0.03 & 0.15 & 0.05 & -0.48 & 0.06 & 2002 Aug  6 & DA & Slitmask \\
\object{NGC 7063:LAWDS  3} & 21:24:07.3 & +36:24:45.6 & 20.89 & 0.06 & 0.15 & 0.10 & -0.52 & 0.11 & 2002 Aug  6 & DA & \\
\object{NGC 7063:LAWDS  4} & 21:24:32.9 & +36:27:51.5 & 21.14 & 0.04 & 0.41 & 0.08 & -0.45 & 0.09 & 2002 Aug  7 & DA & Slitmask \\
\object{NGC 7063:LAWDS  5} & 21:24:43.6 & +36:28:05.6 & 22.73 & 0.13 & 0.43 & 0.23 & -0.37 & 0.28 & 2002 Aug  7 &\nodata\tablenotemark{a}& Slitmask \\
\object{NGC 7063:LAWDS  6} & 21:24:49.4 & +36:33:16.4 & 21.97 & 0.14 & 0.42 & 0.24 & -0.40 & 0.26 & 2002 Aug  6 & DA & \\
\object{NGC 7063:LAWDS 12} & 21:24:33.5 & +36:30:33.5 & 22.49 & 0.14 & 0.39 & 0.24 & -0.60 & 0.28 & 2002 Aug  7 &\nodata\tablenotemark{a}& Slitmask \\
\object{NGC 7063:LAWDS 16} & 21:24:50.4 & +36:34:08.2 & 21.07 & 0.05 & 0.48 & 0.09 & -0.74 & 0.01 & 2002 Aug  6 & QSO& $z=1.965$ \\
\object{NGC 7063:LAWDS 18} & 21:24:20.7 & +36:35:32.0 & 20.06 & 0.06 & 0.27 & 0.10 & -0.46 & 0.11 & 2002 Aug  6 &  A & Slitmask \\
                  &            &             &       &      &      &      &       &      & 2002 Dec  8 &    & \\
\enddata
\tablecomments{Units of right ascension are hours, minutes and seconds, and 
units of declination are degrees, arcminutes, and arcseconds.  Coordinates 
are for Equinox J2000.0}
\tablenotetext{a}{Insufficient signal to identify spectrum.}
\end{deluxetable*}

\section{Spectroscopic Observations \label{sec.specobs}}
Spectroscopic observations were taken between 2001 August and 2005
November using the blue channel of the LRIS spectrograph
\citep{1995PASP..107..375O,1998SPIE.3355...81M} on Keck I (see Table
\ref{tab.obslog}).  The 2001 August observations used the initial
LRIS-B engineering-grade $2048\times 2048$ SITe CCD; all other
observations used the New Blue Camera, consisting of two 2k$\times$4k
Marconi CCDs.  The 2001 observations therefore have lower sensitivity
in the blue.

We selected the 400 groves mm$^{-1}$, 3400\AA-blaze grism, as it is
the available grism with the highest throughput for the vital
higher-order Balmer lines.  The D560 dichroic was used to permit
simultaneous observations of the H$\alpha$ line, though these
observations are not presented here.  Many 2001 observations used a
multi-slit mask with 1\arcsec-wide slitlets; these masks were
typically not at the parallactic angle, and a substantial loss of blue
light is apparent in these spectra.  Subsequent observations used a
1\arcsec ~longslit at parallactic angle. The resulting spectral
resolution (full-width half-max, FWHM) is $\approx 6$\AA.

We reduced the spectra using the \emph{onedspec} package in IRAF.
Overscan regions were used to subtract the amplifier bias, and a
normalized flat field was applied to the data.  Due to very low flux
from the flat field lamps, the quality of the flat-fielding blueward
of $\approx 4000$\AA ~is uncertain.  Cosmic rays were removed from the
two-dimensional spectrum using the ``L.A.Cosmic'' Laplacian cosmic ray
rejection routine \citep{2001PASP..113.1420V}. We then co-added
multiple exposures of individual objects and extracted the
one-dimensional spectrum.  We applied a wavelength solution derived
from Hg, Cd and Zn lamp spectra.  We determined and applied a relative
flux calibration from longslit spectra of multiple spectrophotometric
standard stars.  We made no attempt at obtaining absolute
spectrophotometry for any object.

Spectroscopic identification of each WD candidate is given in Table
\ref{tab.cands.n6633} for NGC 6633 and Table \ref{tab.cands.n7063} for
NGC 7063.  The major non-WD contaminants in the sample are hot
subdwarfs (sdB and sdO stars), A-type stars, and AGN.  We measured the
breadth of H$\delta$ at 20\% below the pseudocontinuum level for a
random sub-sample of the A-type spectra; these were found to match the
criterion for field horizontal branch stars \citep[width $\leq 30$
\AA;][]{1988ApJS...67..461B}.  For QSOs, we determined redshifts by
cross-correlating the spectra with the Sloan Digital Sky Survey
composite QSO spectrum of \citet{2001AJ....122..549V} as described in
\citet{2004AJ....128.1784W}.

\section{Model Fitting and Testing\label{sec.specfits}}
One of the most successful means of measuring DA WD properties is the
simultaneous fitting of the Balmer line profiles in optical spectra, a
method described in \citet{1992ApJ...394..228B} and used in numerous
subsequent works
\citep[e.g.,][]{1994ApJ...432..305B,1995ApJ...444..810B,1997ApJ...488..375F,2001ApJ...563..987C,2005ApJ...618L.123K}.
We adopt this method for analysis of our WD spectra, though we have
made some modifications to the algorithm.  The major difference is
that \citet{1992ApJ...394..228B} use the Levenberg-Marquardt method to
determine the best fit and errors, whereas we use a brute-force
method, considering the entire model atmosphere grid to find the best
fit and a Monte Carlo simulation to determine the errors. We also use
a different grid of model atmospheres.  These and other minor
differences necessitate comparison of the solutions from our routine
and that of \citet{1992ApJ...394..228B}.  In this section, we outline
our fitting technique, our error analysis, and compare our fits with
fits from the literature.

\begin{figure}
\includegraphics[width=3.25in]{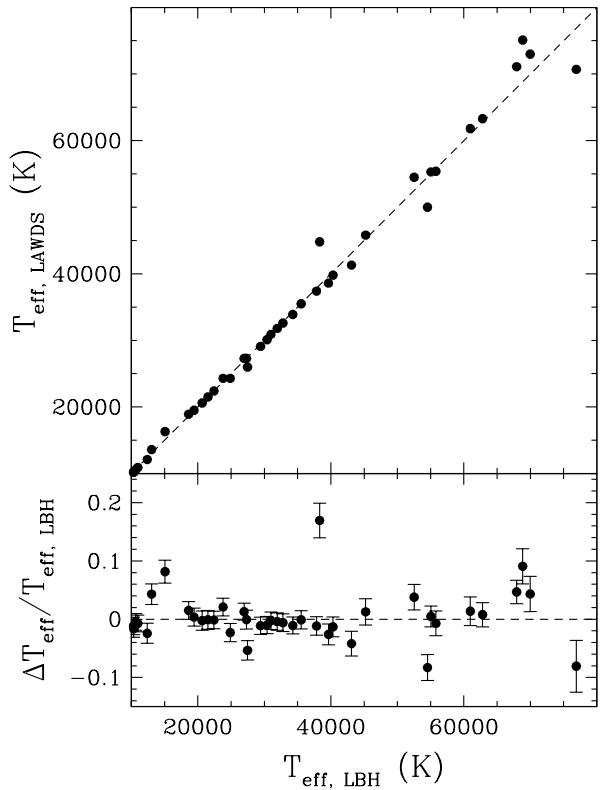}
\caption{Comparison of fit \teff.  Points compare temperatures
  published in \citet[][abscissa]{2005ApJS..156...47L} and those from
  our fits to the same spectra.  \emph{(top)} Absolute \teff ~values;
  the dashed line indicates equivalence.  \emph{(bottom)} Relative
  \teff ~differences.  Error bars are those published in
  \citet{2005ApJS..156...47L}.  The agreement is seen to be excellent.
  \label{fig.pg_teff}}
\end{figure}

\begin{figure}
\includegraphics[width=3.25in]{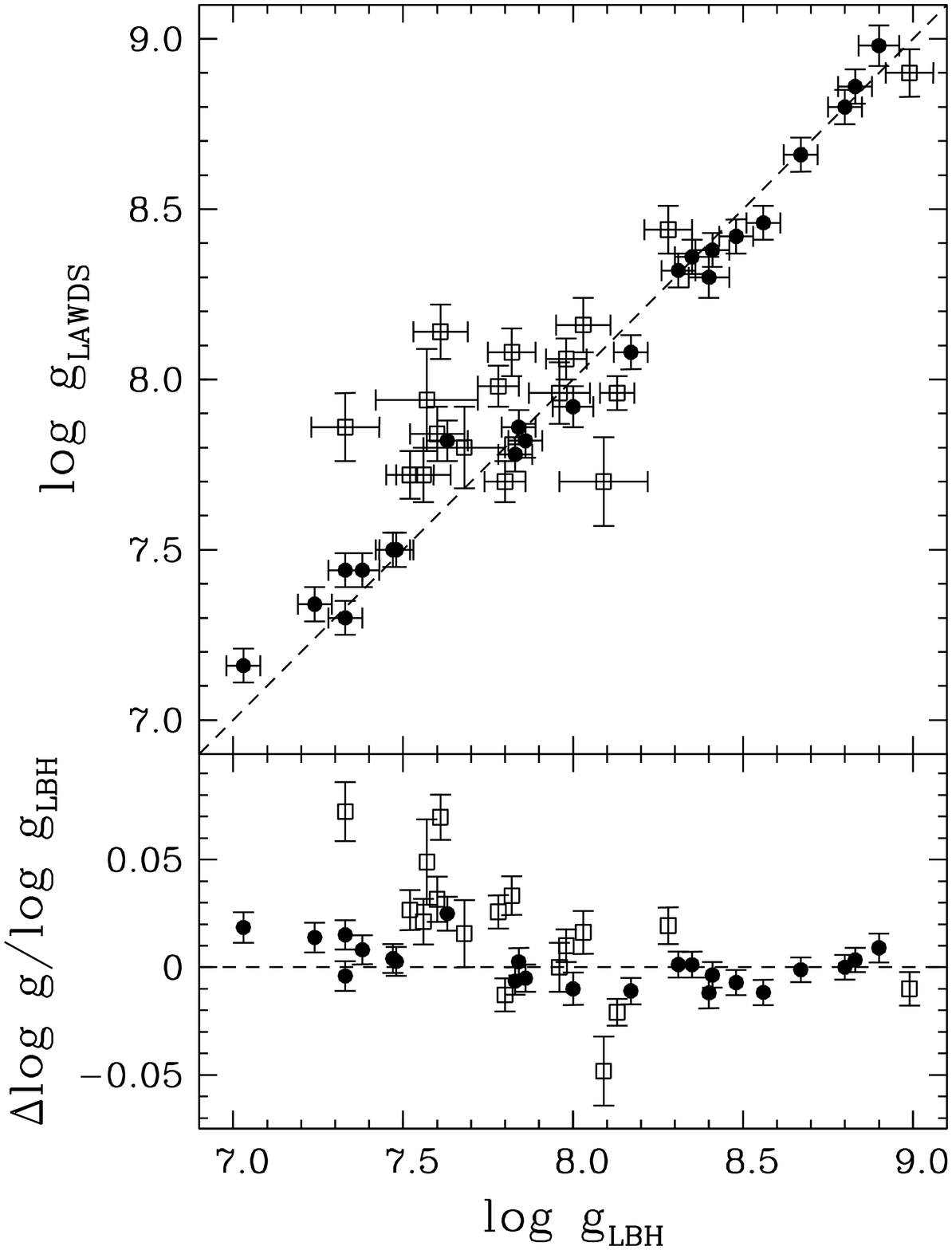}
\caption{Comparison of fit \logg.  Points compare gravities published
  in \citet[][abscissa]{2005ApJS..156...47L} and those from our fits
  to the same spectra.  Very hot ($\teff > 35,000$ K) WDs are shown as
  open symbols; filled symbols indicate WDs with lower temperatures.
  \emph{(top)} Absolute \logg ~values; the dashed line indicates
  equivalence. \emph{(bottom)} Relative \logg ~differences.  Error
  bars are those published in \citet{2005ApJS..156...47L}. For
  $\teff\leq 35,000$ K, the agreement is excellent; our fits to hot,
  low-gravity WDs trend toward higher \logg
  ~values. \label{fig.pg_logg}}
\end{figure}

\subsection{Spectral Fitting Routine}  
The atmospheric models used in the fitting were graciously provided by
D.~Koester, and are slightly modified versions of synthetic, pure-H
atmospheres used in \citet{1997ApJ...488..375F}.  These models cover a
range in \teff ~from $10,000$ to $80,000\,{\rm K}$ and in \logg ~from
7 to 9.  These models are interpolated to create a grid with
$\Delta\teff=100\,{\rm K}$ and $\Delta\logg = 0.02$.  The models are
then convolved with a FWHM$=6$\AA ~Gaussian to match our spectroscopic
instrumental resolution.

\tabletypesize{\footnotesize}
\begin{deluxetable*}{lccccccccc}
\tablewidth{0pt}
\tablecolumns{10}
\tablecaption{Comparison of field white dwarf spectral fits\label{tab.field_comp}}
\tablehead{\colhead{Object}& \colhead{\teff} &
  \colhead{$\sigma_{T_{\rm eff}}$} & \colhead{\logg} &
  \colhead{$\sigma_{\logg}$} & \colhead{\teff,\,\logg} &
  \colhead{\teff,\,\logg} & \colhead{\teff,\,\logg} & \colhead{\teff,\,\logg} &
  \colhead{\teff,\,\logg}\\
  & & & & & \colhead{(1)} & \colhead{(2)} & \colhead{(3)} & 
  \colhead{(4)} & \colhead{(5)} }
\startdata
\object{WD 0001+433} & 41800 &  90 & 8.62 & 0.01 & \nodata      & \nodata      & \nodata       & $46717,8.88$  & \nodata \\
\object{WD 0048+202} & 19400 &  50 & 7.94 & 0.01 & $20340,7.97$ & \nodata      & \nodata       & \nodata       & $20160,7.99$\\
\object{WD 0214+568} & 21000 &  50 & 7.96 & 0.01 & $21180,7.83$ & \nodata      & \nodata       & $21633,7.864$ & \nodata \\
\object{WD 0346-011} & 42400 &  80 & 8.98 & 0.02 & $40540,9.22$ & \nodata      & $43102,9.092$ & \nodata       & \nodata \\
\object{WD 0349+247} & 32000 &  50 & 8.50 & 0.01 & $32920,8.60$ & $31660,8.78$ & \nodata       & \nodata       & \nodata \\
\object{WD 0352+096} & 14800 &  50 & 8.18 & 0.01 & \nodata      & $14770,8.16$ & \nodata       & \nodata       & \nodata \\
\object{WD 0438+108} & 25800 &  50 & 8.04 & 0.01 & $27230,8.05$ & $27390,8.07$ & \nodata       & \nodata       & \nodata \\
\object{WD 0501+524} & 61500 & 220 & 7.58 & 0.01 & \nodata      & $64100,7.69$ & \nodata       & $61193,7.492$ & \nodata \\
\object{WD 0549+158} & 33300 &  70 & 7.72 & 0.01 & \nodata      & \nodata      & $33753,7.664$ & $32747,7.683$ & \nodata \\
\object{WD 0937+505} & 34100 &  30 & 7.76 & 0.01 & \nodata      & \nodata      & \nodata       & $35552,7.762$ & $35830,7.88$ \\
\object{WD 1105-048} & 15700 &  50 & 7.82 & 0.01 & \nodata      & \nodata      & $15576,7.805$ & \nodata       & \nodata \\
\object{WD 1201-001} & 20300 &  50 & 8.26 & 0.01 & $19960,8.26$ & \nodata      & \nodata       & \nodata       & $19770,8.26$ \\
\object{WD 1451+006} & 26000 &  50 & 7.82 & 0.01 & $25670,7.83$ & \nodata      & $26371,7.926$ & $26066,7.704$ & $24930,7.89$ \\
\object{WD 1737+419} & 21900 &  50 & 7.82 & 0.01 & \nodata      & \nodata      & \nodata       & $19458,7.968$ & \nodata \\
\object{WD 1936+327} & 21100 &  50 & 7.86 & 0.01 & $21260,7.84$ & \nodata      & \nodata       & $21240,7.765$ & \nodata \\
\object{WD 2025+554} & 31000 &  50 & 7.82 & 0.01 & \nodata      & \nodata      & \nodata       & $30486,7.762$ & \nodata \\
\object{WD 2046+396} & 65700 & 270 & 7.80 & 0.04 & \nodata      & \nodata      & \nodata       & $63199,7.766$ & \nodata \\
\object{WD 2309+105} & 58800 & 270 & 7.80 & 0.02 & \nodata      & \nodata      & $57990,8.073$ & $58701,7.807$ & $54410,7.90$ \\
\object{WD 2357+296} & 49700 &  70 & 7.74 & 0.01 & \nodata      & \nodata      & \nodata       & $49939,7.596$ & $51960,7.52$ \\
\enddata
\tablerefs{(1) \citealt{1992ApJ...394..228B}, (2) \citealt{1995ApJ...444..810B},
(3) \citealt{1995ApJ...443..735B}, (4) \citealt{1997ApJ...488..375F}, 
(5) \citealt{2005ApJS..156...47L}}
\end{deluxetable*}

For the line fitting, Balmer-series lines from H$\beta$ through H9 are
considered.  For both the model and observed spectra, a linear fit is
made to the pseudo-continuum on either side of the line, and the
spectra are normalized to the linear fit.  The model flux in each
pixel is determined by averaging the flux at ten equally-spaced
wavelengths in the pixel; the noise is measured empirically from the
spectral regions used to fit the continuum. The $\chi^2$ fit of the
model to the observations is then calculated for each pixel in the
one-dimensional extracted spectrum.  A separate $\chi^2$ value is
calculated for each Balmer line $i$ at each \teff ~and \logg ~in the
grid.  A global minimum $\chi^2_{i,min}$ is then found for each Balmer
line.

For $\Delta_i (\teff,\logg)\equiv \chi^2_i(\teff, \logg) -
\chi^2_{i,min}$, the probability $p_i$ that a $\chi^2_i$ variable
exists with a smaller $\Delta_i$ is
\begin{equation}
p_i = 1 - \Gamma(\frac{\nu}{2},\frac{\Delta_i}{2})\label{eqn.gamma}
\end{equation}
where $\Gamma$ is the incomplete gamma function and $\nu\,(=2)$ is the
number of degrees of freedom \citep{Recipes}.  In other words,
Eq.~\ref{eqn.gamma} gives the probability that the best-fit exists
within a contour of constant $\Delta_i$.

The combined probability $1-P$ that the best-fit model lies
\emph{outside} a given contour of $\Delta (\teff,\logg)$ is then:
\begin{equation}
1-P(\teff,\logg) = \prod_i 1-p_i = \prod_i \Gamma(\frac{\nu}{2},\frac{\Delta_i}{2})\,.
\end{equation}
The global minimum of $P(\teff,\logg)$ then gives the best-fitting
\teff ~and \logg.

In order to reduce the computational time, we use an adaptive grid
spacing.  Starting with a course grid spacing, we determine the global
best fit and recenter a smaller and finer grid on that fit.  We
iterate this process until we reach our model grid resolution of
$\Delta \teff=100{\rm K}$ and $\Delta\logg=0.02$.

For objects with very high signal-to-noise ratios (S/N$\gtrsim 240$
per resolution element), numerical errors prevent determination of
\teff ~and \logg.  The reasons for this failure are small systematics
in flat-fielding and/or continuum fitting; due to the very high
signal-to-noise, even these small deviations result in very large
$\chi^2$ values for which $P$ cannot be calculated due to
computational numerical limitations.  This was readily solved by
setting a minimum noise value, such that
$\sigma_\lambda/f_{\lambda,{\rm obs}}\geq 0.004$.

To convert from \teff ~and \logg ~to the WD mass (\mf) and cooling age
(\tcool), we used evolutionary models provided by P.~Bergeron.  These
models include synthetic photometry and cooling ages for WDs with CO
cores and thick hydrogen layers ($M_H/M_* = 10^{-4}$), using WD
evolutionary models from \citet{1995LNP...443...41W} for $\teff\geq
30,000{\rm K}$ and from \citet{2001PASP..113..409F} for cooler \teff.
We only use those models for WDs with $0.4\msun \leq M_{\rm WD} \leq
1.2\msun$; although lower-mass carbon-oxygen models are available, any
such low-mass WDs are likely He-core products of binary star
evolution, so use of low mass carbon-oxygen models would be improper.
\mf ~and \tcool ~are interpolated from these models using the
best-fitting \teff ~and \logg.  We also calculate synthetic
photometric indices for each WD by interpolation from the evolutionary
models.

\subsection{Spectral Fitting Error Determination}
As the errors in the calculated \teff ~and \logg ~are correlated, we
choose to determine random errors via Monte Carlo methods.  By design,
such methods also account for potential error sources not included in
the $\chi^2$ fitting, such as the decreased S/N in the absorption line
profiles.

We begin by convolving the best-fit spectral model with the
instrumental resolution.  The convolved model is then multiplied by
the spectroscopic instrument response and scaled such that the number
of counts in a region of the scaled model surrounding H$\beta$ is the
square of the pixel-to-pixel S/N measured in the same region in the
observed spectrum.  Poisson noise is then added to each pixel.
Finally, the instrumental response is divided out and the noisy model
spectrum is fit using our fitting routine.

Nine independent simulations are run for each observed white dwarf,
and the standard deviations about the mean \teff, \logg, \mf, and
\tcool ~are calculated; these standard deviations are the errors
quoted in all fits.  The number of simulations was selected to
maximize the number of points used in the statistical analysis while
minimizing computing time ($\sim \frac{4}{3}$ hr per simulation).
\citet{1990AJ....100...32B} note that, for sample sizes similar to our
nine simulations, the standard deviation does not always give a good
estimate of the scale of the scatter; they find that the ``gapper''
method of \citet{Wainer1976} best recovers the true scatter.  We
calculate the scatter of \teff ~and \logg ~using the gapper method and
find identical values to within our quoted precision.  We therefore
claim that the nine simulations are sufficient in recovering the
measurement errors of \teff ~and \logg.  As the smallest model
spectral grid has a step size of $\Delta\teff=100$ K and $\Delta\logg
= 0.02$, we adopt minimum errors of $\sigma_{T_{\rm eff}}=50$ K and
$\sigma_{\logg}=0.01$.

\subsection{Comparison to Published WD Atmospheric
  Parameters\label{sec.comp_fits}} 
In order to further test our Balmer line fitting procedures and derive
atmospheric parameters for WDs, we make two comparisons of our results
with those in the literature.  First, we fit spectra of WDs from the
Palomar Green (PG) Survey study of \citet{2005ApJS..156...47L}
provided by J.~Liebert.  This provides a direct comparison of the
output from our fitting routine and that of
\citet{1992ApJ...394..228B}.  This fitting also provides a direct
comparison of the atmospheric models of P.\ Bergeron and collaborators
used in \citet{2005ApJS..156...47L} and the atmospheric models of D.\
Koester and collaborators used in our fitting.

Rather than attempt to fit all 348 DA WDs in the PG sample, we
selected a sub-sample of 50 objects, chosen to cover the range of our
model grid in \teff ~and \logg.  This sample was then pared to exclude
three known magnetic WDs and two WDs with composite spectra.  Two hot
WDs were also removed due to the presence of \ion{He}{2} lines,
indicating these are DAOs.  After these cuts, 44 objects remained in
the sample.

Figure \ref{fig.pg_teff} compares the \teff ~from our fits with those
of \citet{2005ApJS..156...47L}.  With few exceptions, agreement is
excellent.  At high temperatures ($\teff\gtrsim 60,000{\rm K}$), our
fit temperatures appear systematically higher than the published
values.  This systematic is not of great concern, as these
temperatures are known to be biased due to the presence of significant
metal opacity in many hot WDs \citep[e.g.,][]{1998A&A...329.1045W},
and because these objects cool so rapidly that even a large fractional
error in \tcool ~leads to a small fractional error in the progenitor
star's age, resulting in only small errors in the derived initial
mass.  The most discrepant \teff~ measurement at cooler \teff~ is for
\object{PG1255+426} ($\Delta\teff=6800{\rm K}$; $\sigma_{\teff} =
1135{\rm K}$); our fit is poor due to the low S/N of the
observation. Excluding the hot WDs and PG1255+426, the standard
deviation in the offsets between the published \teff~ and our fit
\teff~ is $\sigma(\Delta\teff) = 685\,{\rm K}$.

A comparison of the best-fitting surface gravities (Figure
\ref{fig.pg_logg}) reveals a larger scatter than is seen in the
temperatures [$\sigma(\Delta\logg) = 0.17$], though the overall
agreement is good.  There is some evidence for a systematic
discrepancy for hot WDs with $\logg\lesssim 7.75$.  This is due to the
weakness at high \teff ~of the higher-order Balmer lines, which
provide the most leverage for surface gravity determinations.
Excluding WDs with $\teff > 35000\,{\rm K}$, the standard deviation in
the difference of the \logg ~determinations is 0.08.

In summary, our fitting method and model atmospheres recover very
similar atmospheric parameters as the method and atmospheres used in
\citet{2005ApJS..156...47L}, with the exception of the lower gravity
($\logg\lesssim 7.75$), hot ($\teff\gtrsim 35,000{\rm K}$) WDs.  This
agreement gives us confidence that our fitting routine is working as
expected.

We now compare fits of bright WD spectra obtained during our
spectroscopic runs with published atmospheric parameters of these
well-studied objects.  This comparison identifies potential systematic
effects introduced by differing instruments and data reduction
techniques.

We obtained spectra of 19 WDs, each with atmospheric parameters
published in one or more of five studies:
\citet{1992ApJ...394..228B,1995ApJ...444..810B,1995ApJ...443..735B,1997ApJ...488..375F,2005ApJS..156...47L}.
Our fits and those from the literature are given in Table
\ref{tab.field_comp}.  We noted no obvious systematic in temperature
determinations as compared with previous studies, though the scatter
[$\sigma(\Delta\teff) = 1100\,{\rm K}$] is significantly larger than
the quoted internal errors.  There is a systematic offset in surface
gravity for $\logg\gtrsim 8.5$, with our fits having significantly
lower values than previous studies.  However, only a few high-gravity
WDs were investigated, and two of these WDs have published $\logg >
9.0$, the highest gravity in our model grid.  The scatter in the
\logg~ determinations is found to be $\sigma(\Delta\logg) = 0.12$.

\section{Discussion\label{sec.disc}}

\subsection{Candidate White Dwarf Spectral Fits}

\begin{figure}
\includegraphics[width=3.25in]{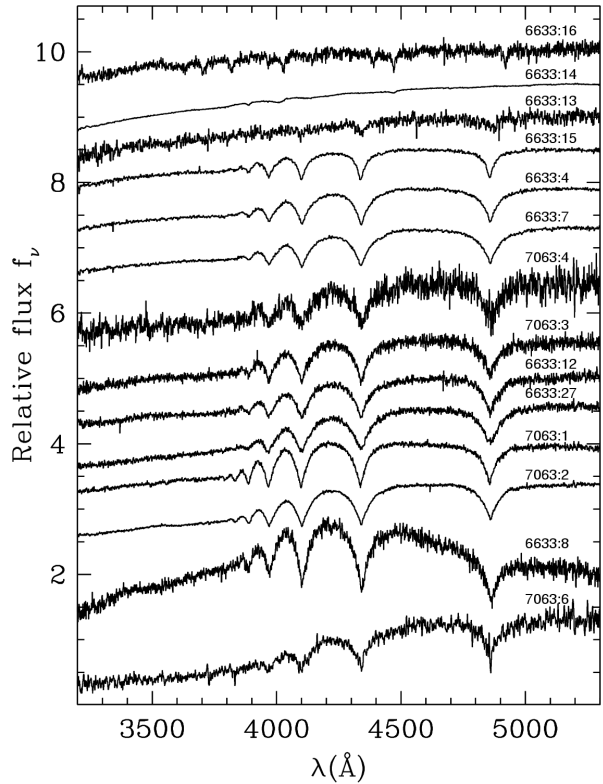}
\caption{Spectra of WDs in the field of NGC 6633 and NGC 7063.  The
  top two spectra are DB WDs; the bottom WD spectrum was unable to be
  fit due to poor sky subtraction in the Balmer line cores.  All other
  spectra are arranged in order of increasing fit \teff (bottom to
  top). Spectra are normalized to unity at 5300\AA ~and offset by
  arbitrary amounts.  No correction for interstellar reddening has
  been applied. Labels are in the format NGC:ID; e.g., 6633:16 = NGC
  6633:LAWDS 16. \label{fig.wd_spec}}
\end{figure}

\begin{figure}
\plotone{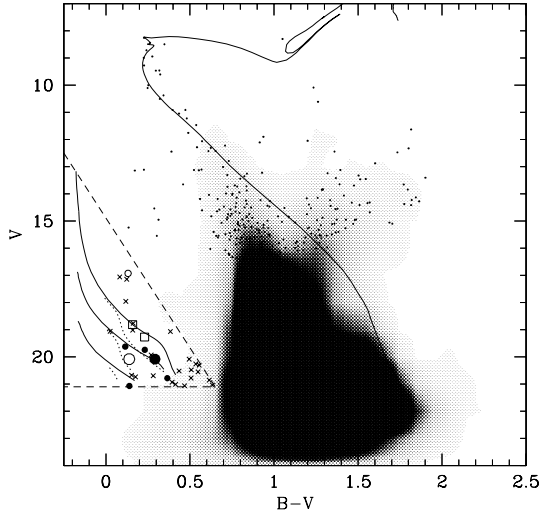}
\caption{Color-magnitude diagram for NGC 6633 with spectroscopic
  identifications.  Grayscale reflects the relative stellar density in
  the CFHT imaging, while points are all stars in the Nickel fields.
  The $\log t=8.75$ isochrone is shown (solid), along with cooling
  curves for for cluster DAs (solid) and DBs (dashed) are given for
  $\log \tcool\leq 8.8$ and masses of 0.4\msun (top), 0.8\msun
  (middle), and 1.2\msun (bottom). The photometric selection criteria
  are dashed lines.  Crosses indicate objects identified as non-WD or
  objects not spectroscopically identified.  Small circles indicate
  non-cluster member DAs (filled) and DBs (open).  The large filled
  circle is the cluster DA, while the large open circle is the
  potential cluster DB.  The open squares are the two potential double
  degenerates.
  \label{fig.n6633_ids}}
\end{figure}

\begin{figure*}
\includegraphics[angle=270,scale=0.6]{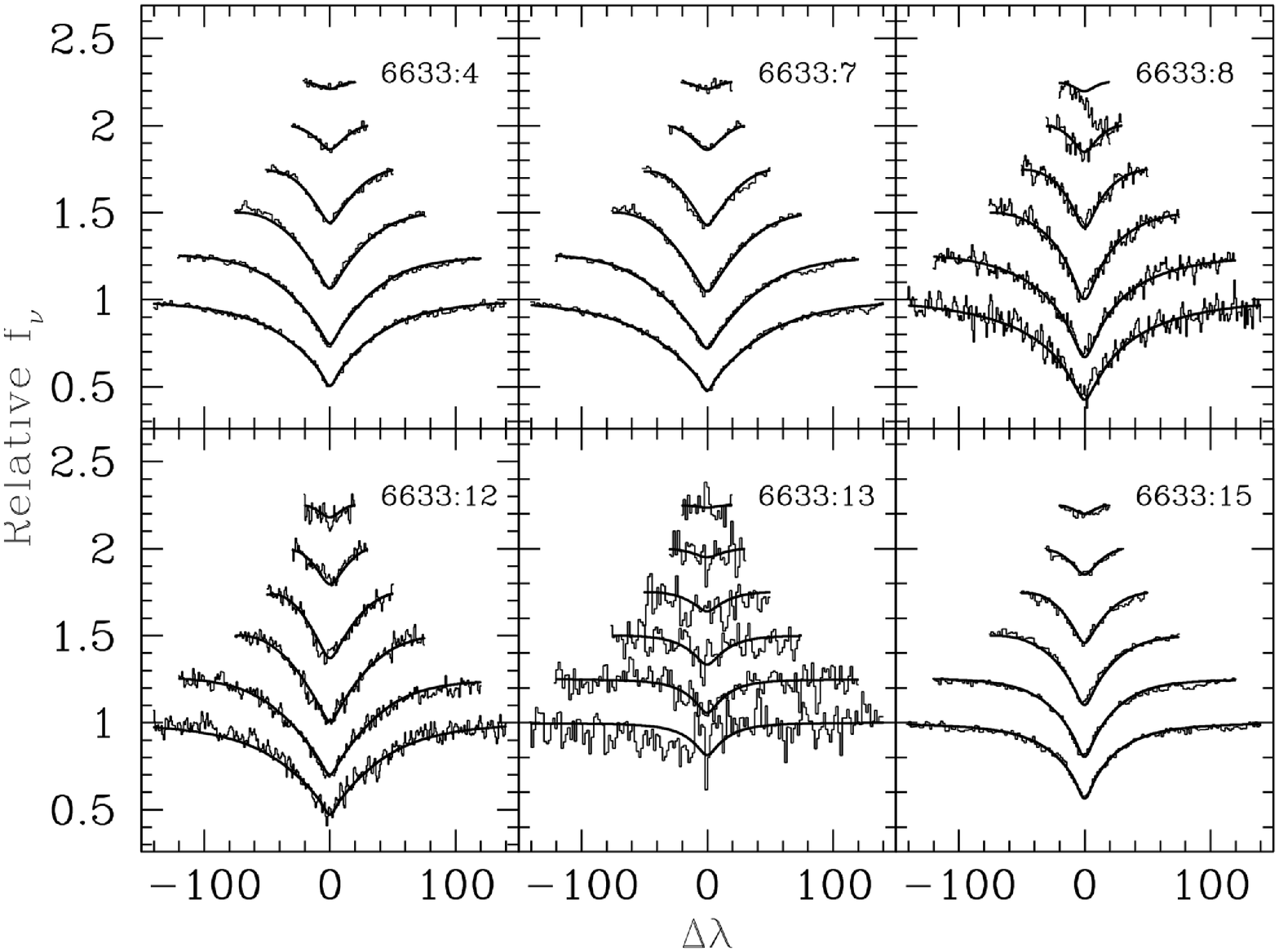}
\caption{Balmer line profiles for all DA WDs in the cluster fields.
  The plotted lines range from H$\beta$ (bottom) to H9 (top), normalized
  to the pseudo-continuum just outside the plotted regions and
  arbitrarily offset vertically.  Thick solid lines are the
  best-fitting models, while the thinner histogram is the observed
  spectrum. Labels are in the format NGC:ID; e.g. 6633:4 = NGC
  6633:LAWDS 4.\label{fig.wd_fits}}
\end{figure*}

Spectra of the WDs in both cluster fields are shown in Figure
\ref{fig.wd_spec}, and each object is also identified in the cluster
CMD (Figure \ref{fig.n6633_ids}).  Using our spectral fitting
technique, we fit the spectra of all DA WDs detected with LRIS.  We
present the results of these fits in Table \ref{tab.wdfits} and show
the Balmer line profiles in Figure \ref{fig.wd_fits}.  We then
interpolate using the evolutionary models of
\citet{2001PASP..113..409F} and photometric models of
\citet{Holberg06} to determine WD masses and cooling ages.  We note a
persistent error in continuum fitting of the H9 line in NGC 6633:LAWDS
8 (visible in Figure \ref{fig.wd_fits}); the H9 line is excluded from
that fit.

\begin{figure*}
\includegraphics[angle=270,scale=0.6]{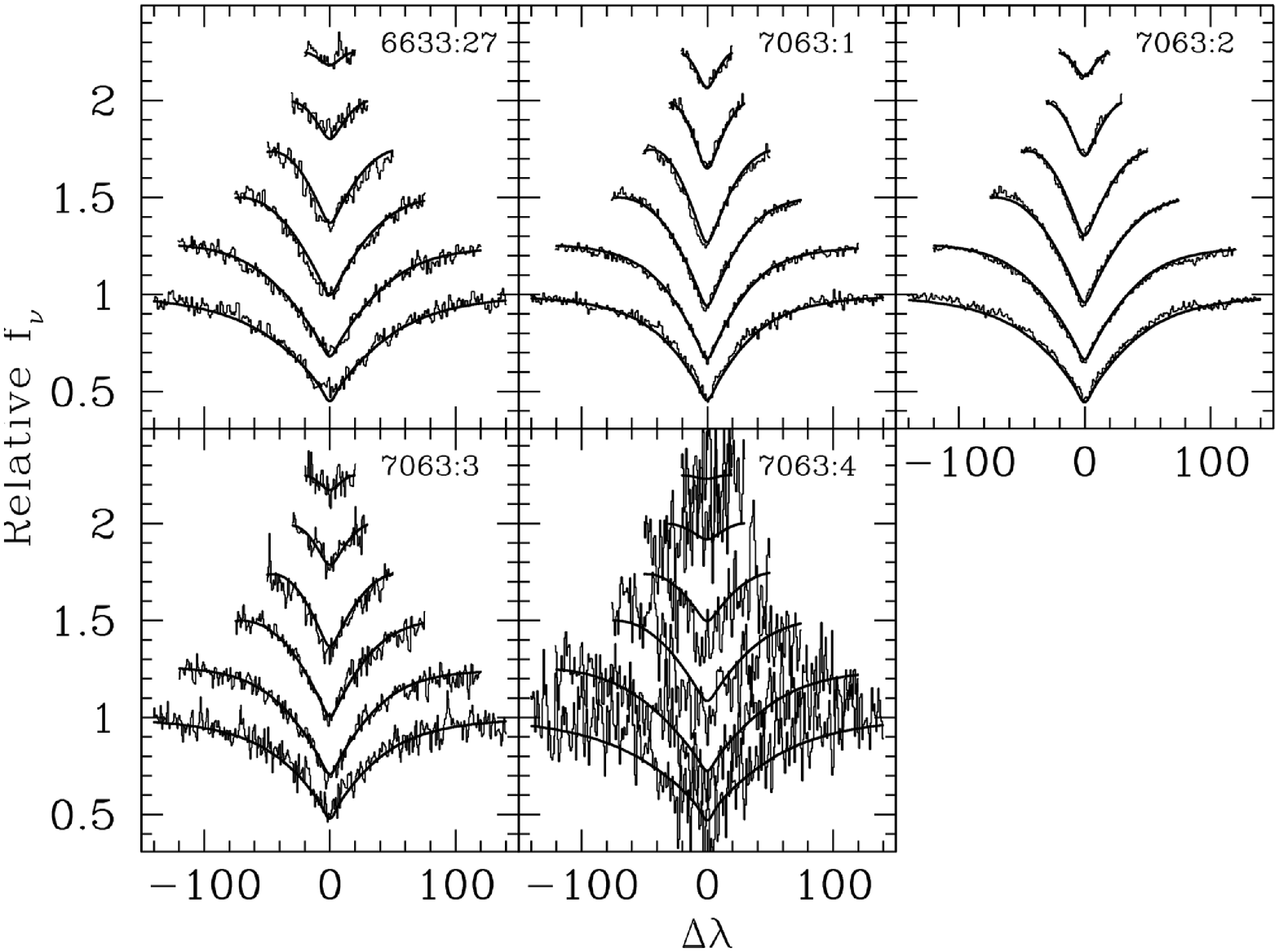}
\centerline{Figure \ref{fig.wd_fits}, cont.}
\end{figure*}

\tabletypesize{\tiny}
\begin{deluxetable*}{llccccccccccc}
\tablecolumns{13}
\tablewidth{0pt}
\tablecaption{DA White Dwarf Spectral Fits\label{tab.wdfits}}
\tablehead{\colhead{Object} & \colhead{Designation\tablenotemark{a}} & \colhead{S/N\tablenotemark{b}} & \colhead{\teff} 
   & \colhead{$\sigma_{\teff}$} & \colhead{\logg} & \colhead{$\sigma_{\logg}$} & \colhead{\mwd} & \colhead{$\sigma_{\mwd}$}
   & \colhead{\tcool} & \colhead{$\sigma_{\tcool}$} & \colhead{$(m-M)_V$} & \colhead{$\sigma_{(m-M)_V}$} \\
   & & & \colhead{(K)} & \colhead{(K)} & & & \colhead{\msun} & \colhead{\msun} & \colhead{$\log({\rm yr})$} & 
   \colhead{$\log({\rm yr})$} & &}
\startdata
NGC 6633:LAWDS  4 & WD J1827+064   & 114 & 20800 &  50 & 8.26 & 0.01 & 0.79 & 0.01 & 8.035 & 0.010 & 7.75 & 0.02 \\
NGC 6633:LAWDS  7 & WD J1827+063.1 & 114 & 18500 &  50 & 8.40 & 0.02 & 0.87 & 0.01 & 8.314 & 0.013 & 7.78 & 0.03 \\
NGC 6633:LAWDS  8 & WD J1827+063.2 &  62 & 11400 &  50 & 8.84 & 0.03 & 1.11 & 0.02 & 9.255 & 0.012 & 6.55 & 0.06 \\
NGC 6633:LAWDS 12 & WD J1828+063   &  97 & 17500 &  90 & 8.12 & 0.02 & 0.69 & 0.01 & 8.168 & 0.017 & 9.64 & 0.04 \\
NGC 6633:LAWDS 13 & WD J1827+063.3 &  34 & 50000 &1110 & 7.24 & 0.16 & 0.43 & 0.04 & 5.674 & 0.388 &13.29 & 0.31 \\
NGC 6633:LAWDS 15 & WD J1826+064   & 167 & 27200 &  50 & 7.72 & 0.01 & 0.49 & 0.01 & 7.098 & 0.003 & 9.90 & 0.02 \\
NGC 6633:LAWDS 27 & WD J1827+063.4 &  79 & 16100 &  90 & 8.24 & 0.02 & 0.77 & 0.01 & 8.370 & 0.015 & 8.61 & 0.04 \\
NGC 7063:LAWDS  1 & WD J2124+363.1 & 113 & 15500 &  60 & 7.50 & 0.01 & 0.37 & 0.01 & 7.930 & 0.008 & 9.61 & 0.04 \\
NGC 7063:LAWDS  2 & WD J2124+363.2 & 162 & 15300 &  50 & 7.86 & 0.01 & 0.53 & 0.01 & 8.169 & 0.008 & 8.20 & 0.03 \\
NGC 7063:LAWDS  3 & WD J2124+364.1 & 111 & 17900 & 140 & 8.04 & 0.03 & 0.65 & 0.02 & 8.076 & 0.028 & 9.90 & 0.08 \\
NGC 7063:LAWDS  4 & WD J2124+364.2 &  22 & 18300 & 190 & 8.80 & 0.06 & 1.09 & 0.03 & 8.629 & 0.053 & 8.90 & 0.13 \\
\enddata
\tablenotetext{a}{\citet{1999ApJS..121....1M}-style designation}
\tablenotetext{b}{Average signal-to-noise per resolution element at pseudo-continuum surrounding H$\delta$}
\end{deluxetable*}

\begin{figure*}
\begin{minipage}{3.25in}
\includegraphics[scale=0.45]{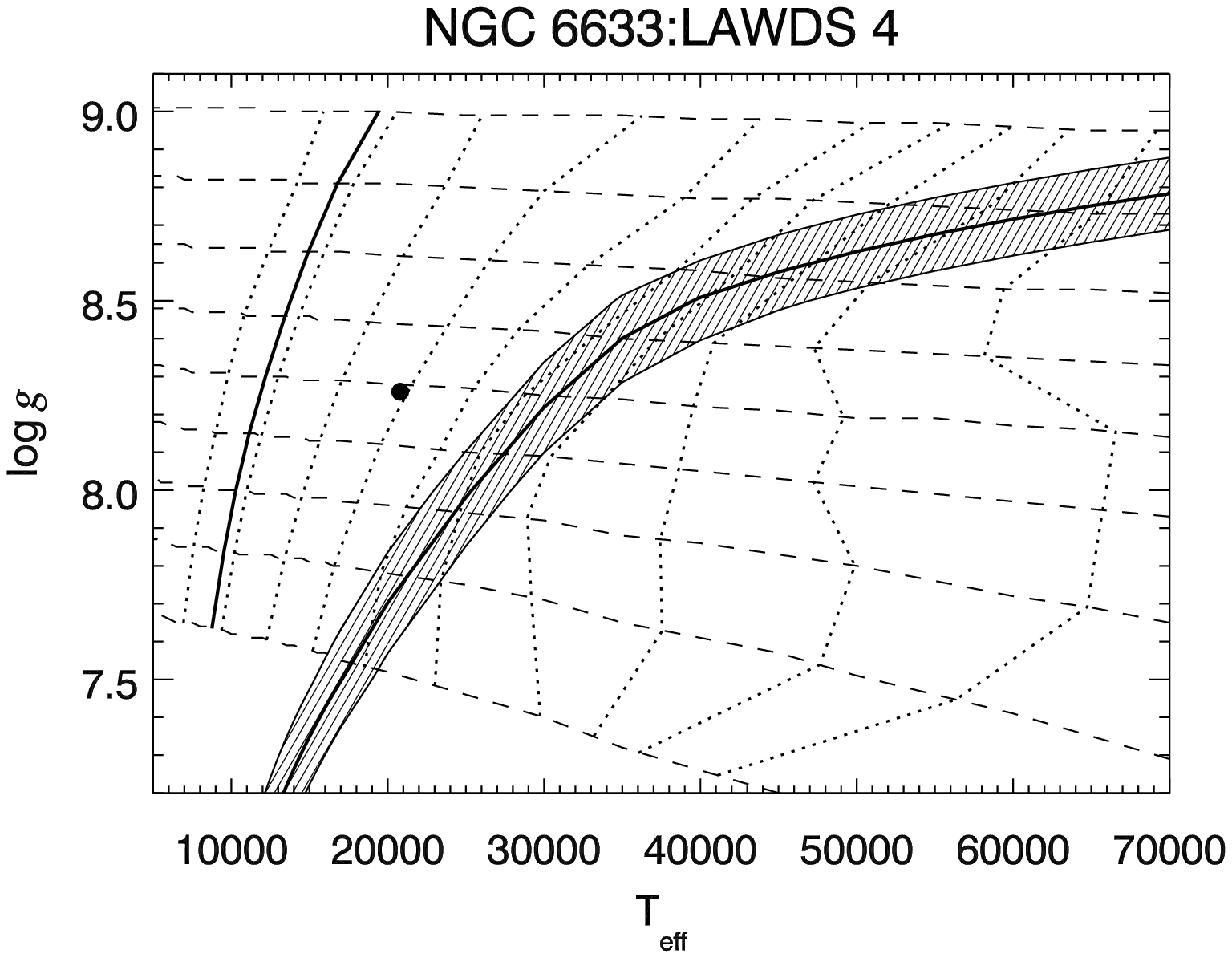}
\includegraphics[scale=0.45]{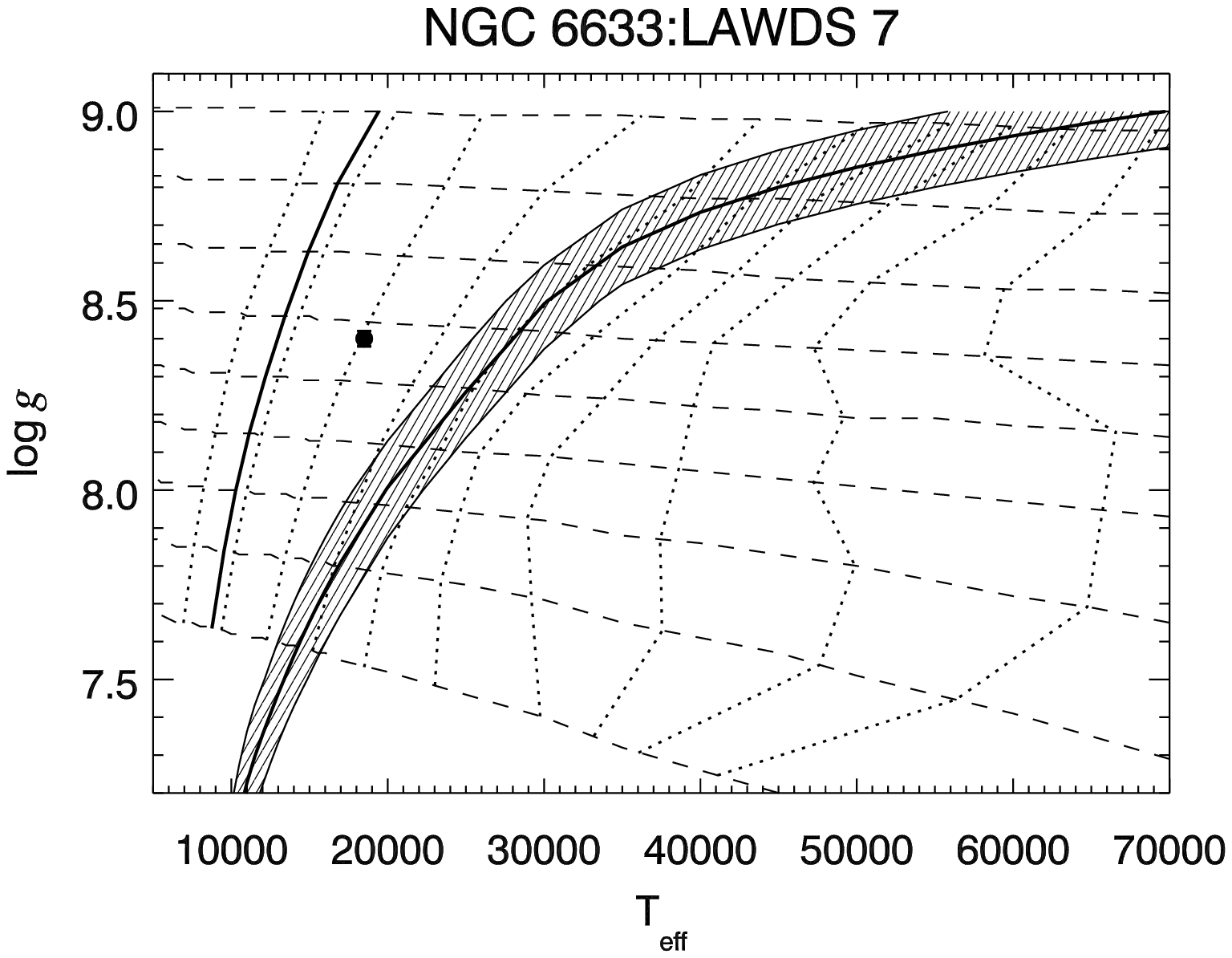}
\includegraphics[scale=0.45]{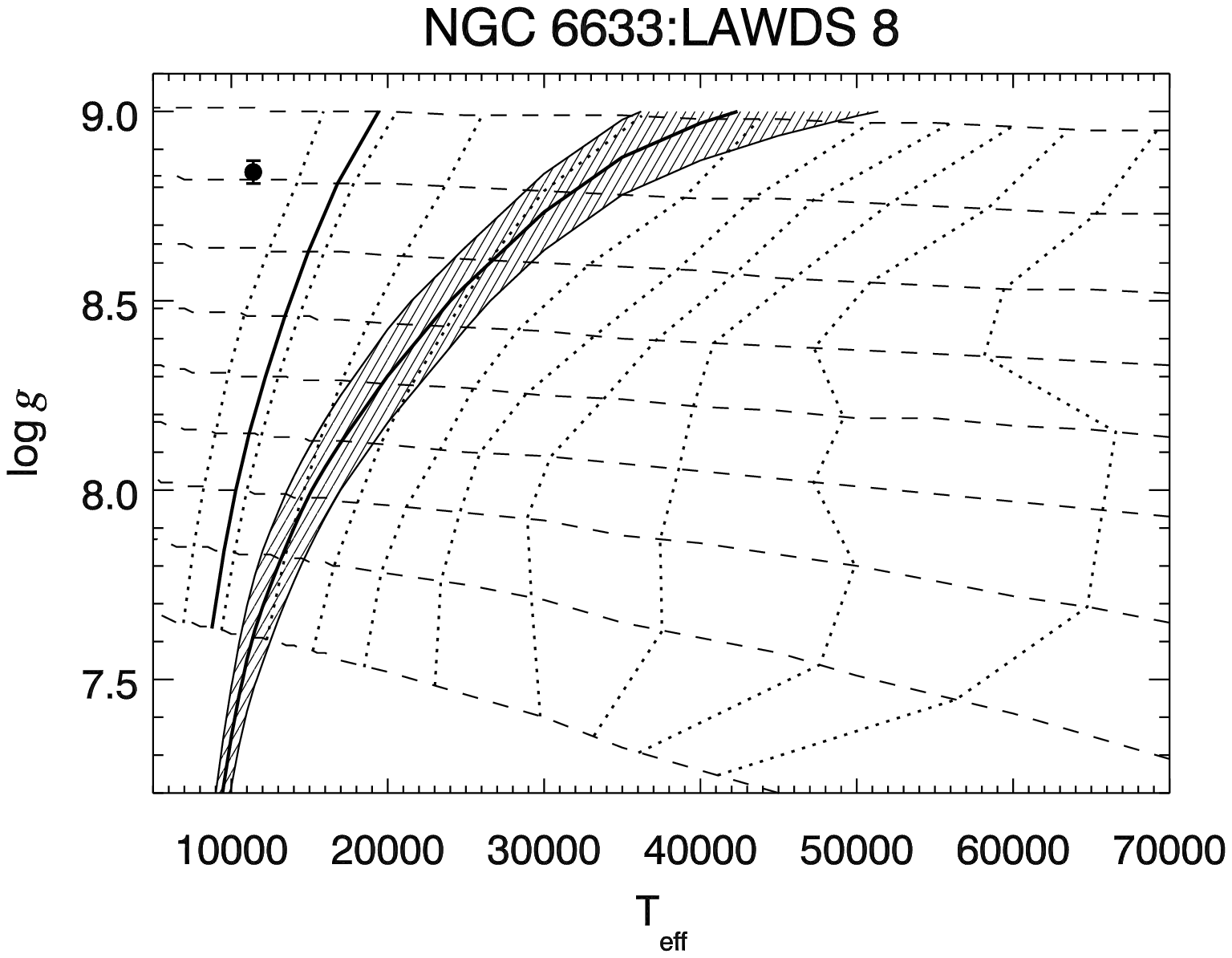}
\includegraphics[scale=0.45]{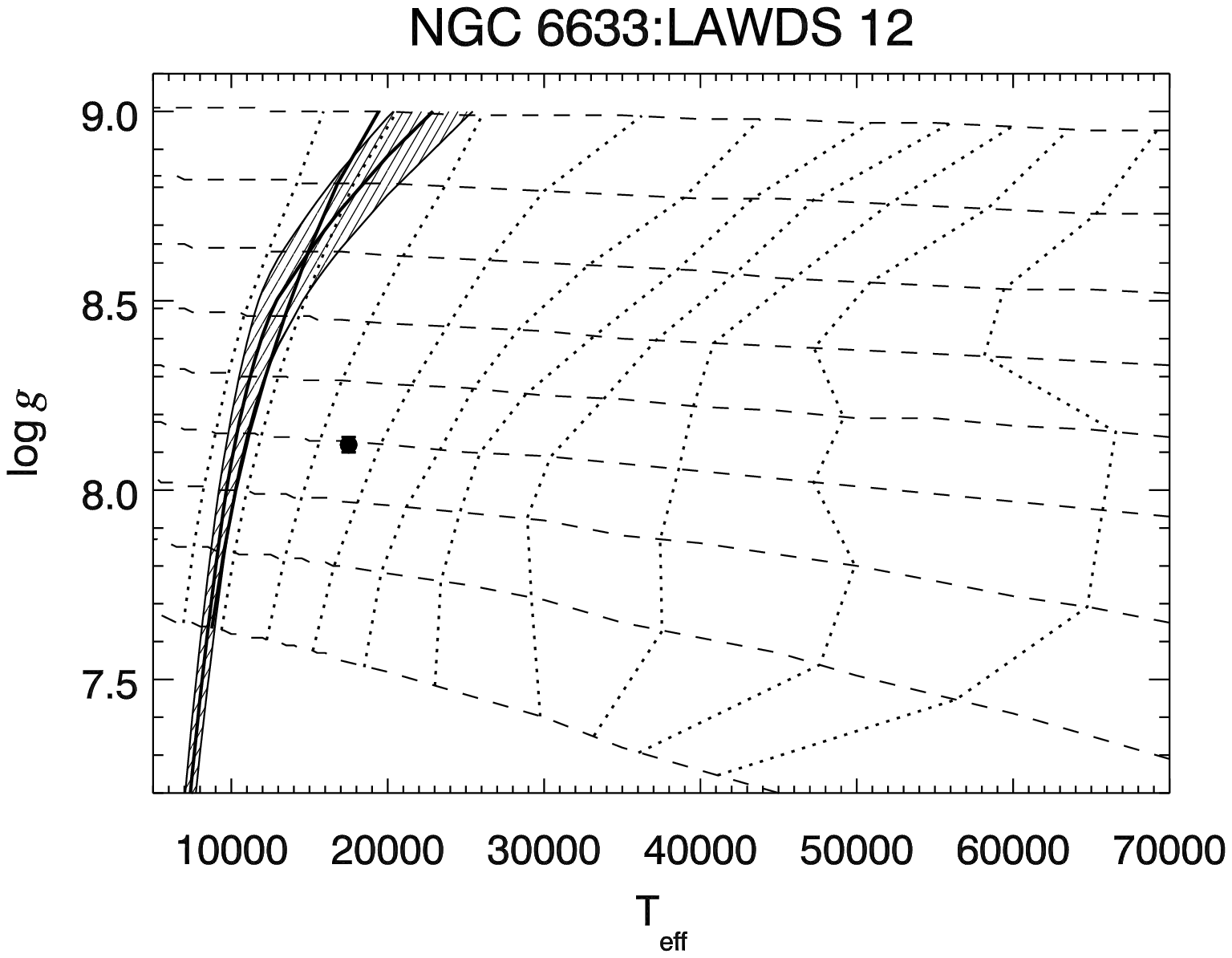}
\end{minipage}\hfill
\begin{minipage}{3.25in}
\includegraphics[scale=0.45]{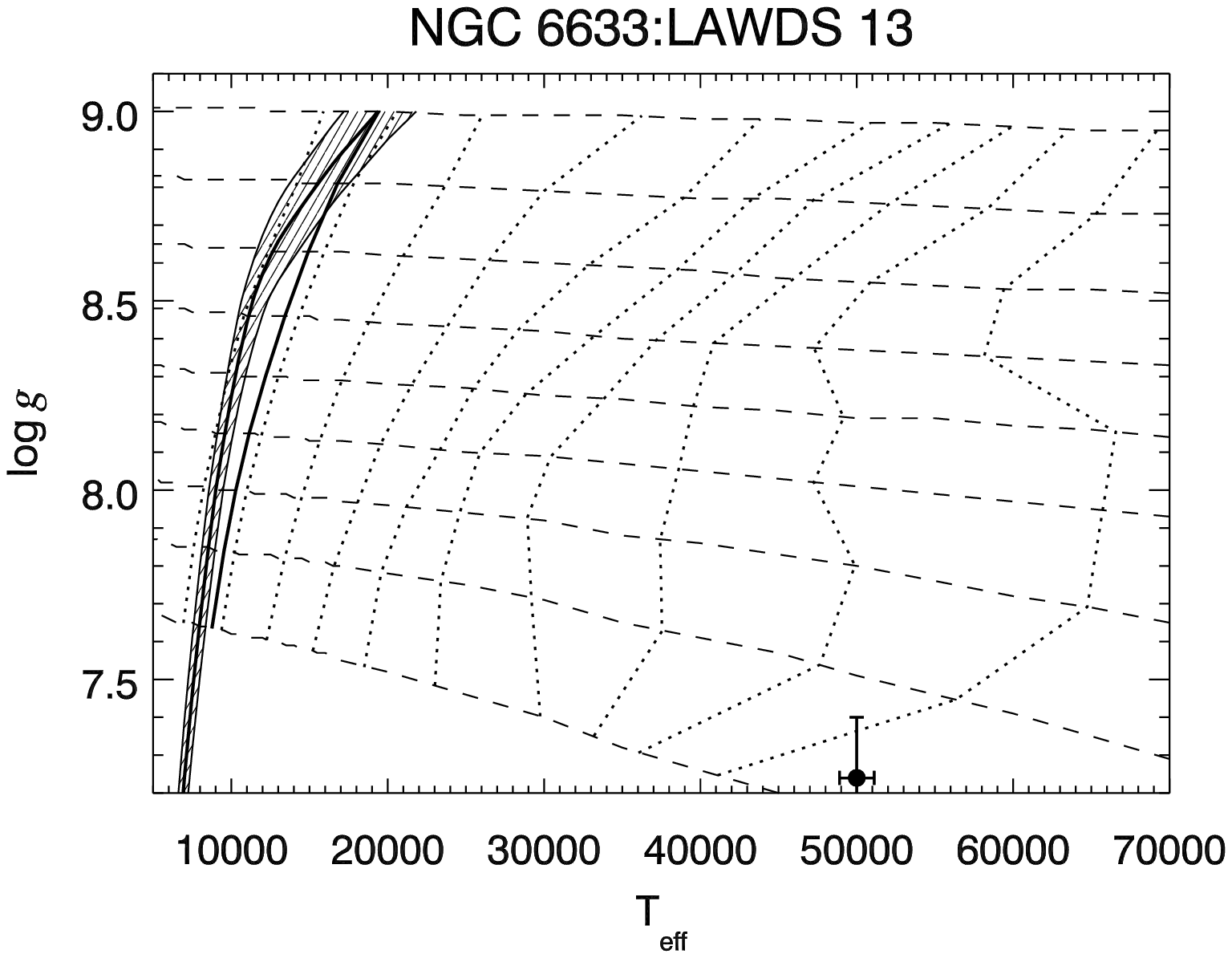}
\includegraphics[scale=0.45]{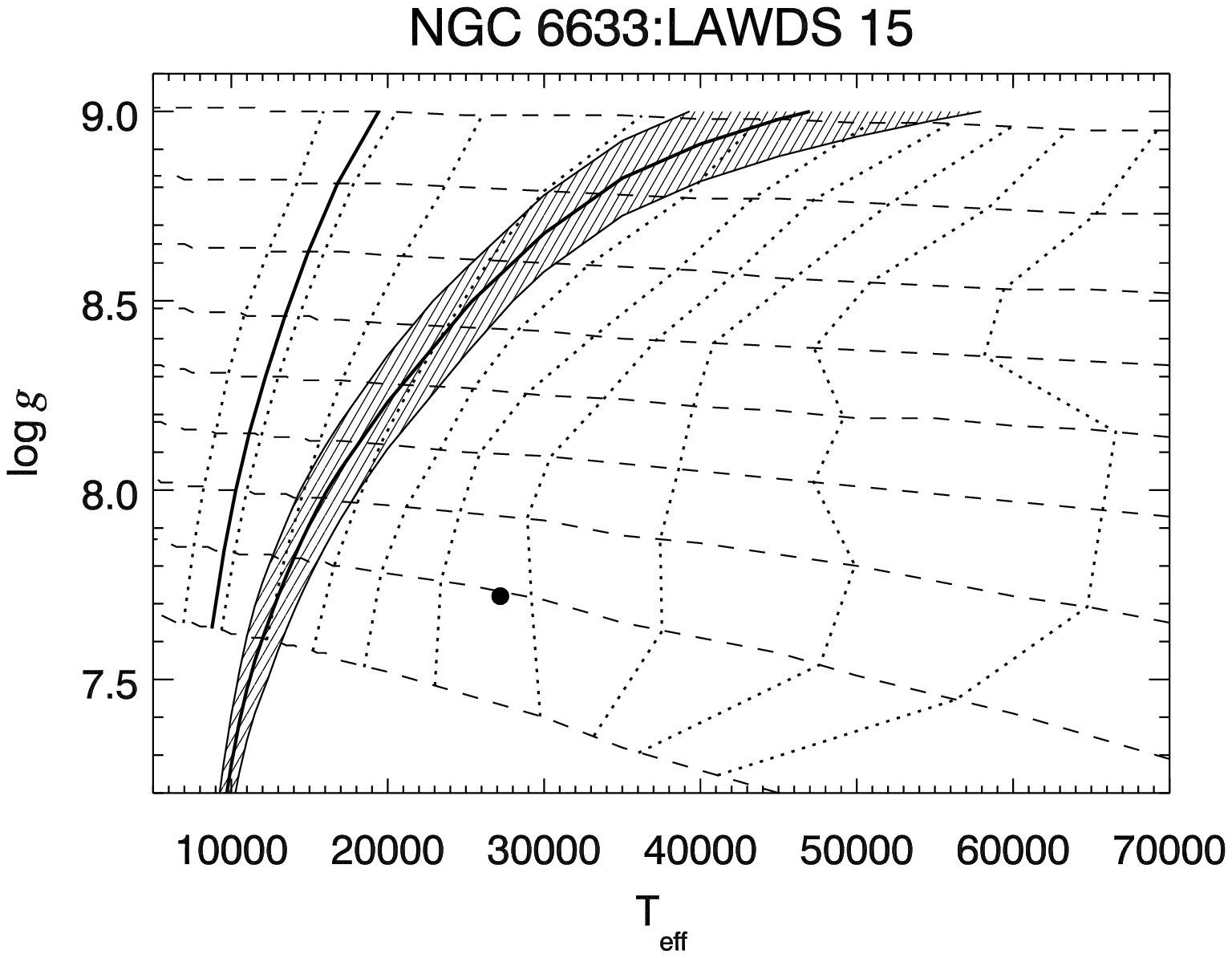}
\includegraphics[scale=0.45]{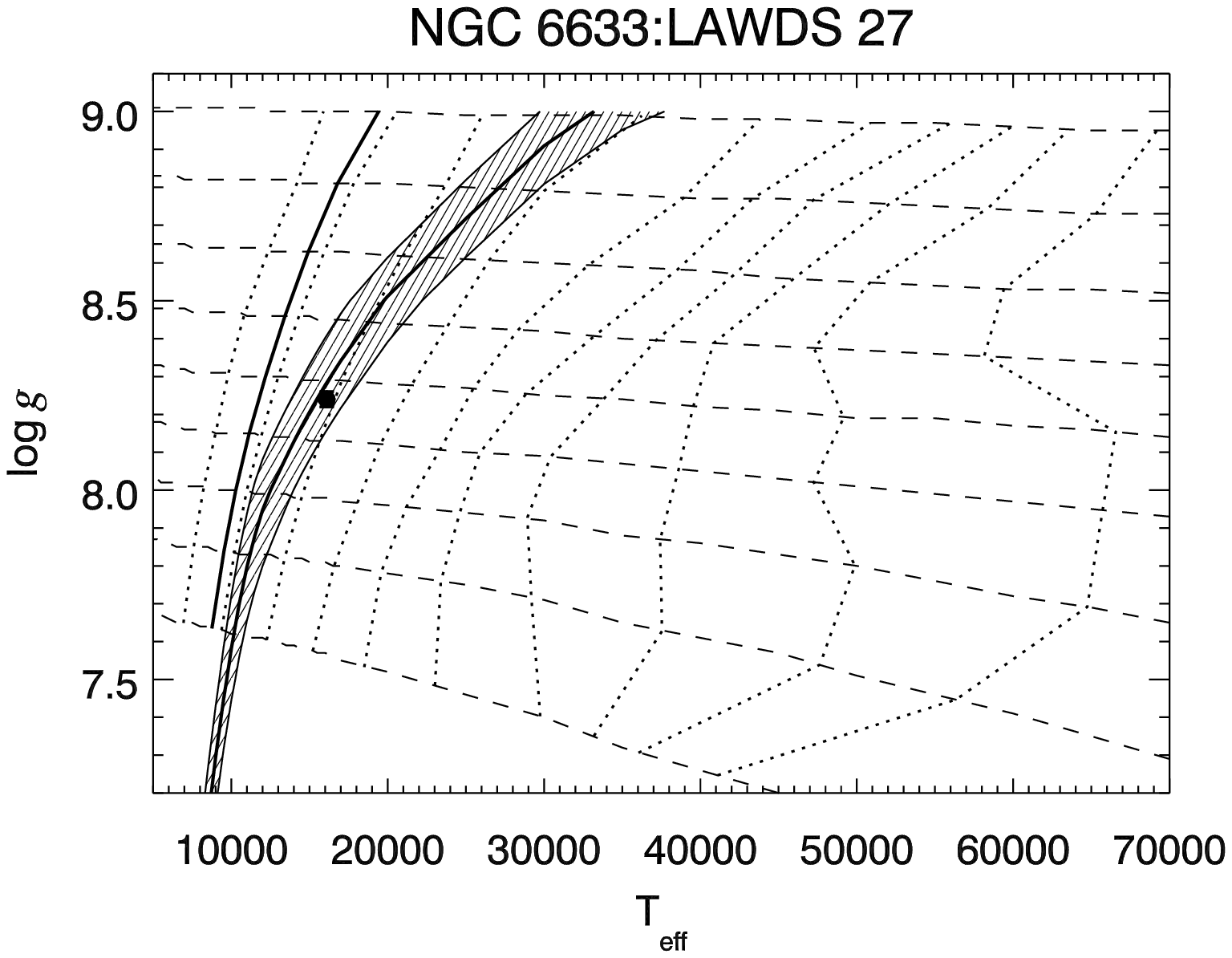}
\end{minipage}

\caption{Physical parameters and photometric distance modulli for WDs
  in the field of NGC 6633.  Points with 1$\sigma$ error bars (often
  smaller than the points) indicate our spectral fits.  Horizontal,
  dashed lines are cooling tracks for WDs with masses of 0.4\msun
  (bottom) to 1.2\msun (top) in 0.1\msun intervals.  Nearly-vertical,
  dotted lines are lines of constant WD cooling ages from 1 Myr
  (right) to 1 Gyr (left) in logarithmic intervals of 0.33 dex. The
  heavy, nearly-vertical curve indicates where WD cooling ages
  equal the cluster age; WDs to the left of this line are too old
  to be cluster members.  The shaded region indicates the region of
  \teff, \logg ~space where cluster white dwarfs would have
  photometric indices consistent at the $2\sigma$ level with the
  observed WD's photometry.   This region includes
  errors in the cluster distance modulus, foreground extinction and
  the observed $V$-band photometric error. If a WD lies in the shaded
  region, it is photometrically consistent with cluster membership; if
  it lies outside this region, it is inconsistent with cluster membership.
  \label{fig.photdist_n6633}}
\end{figure*}

\begin{figure*}
\begin{minipage}{3.25in}
\includegraphics[scale=0.45]{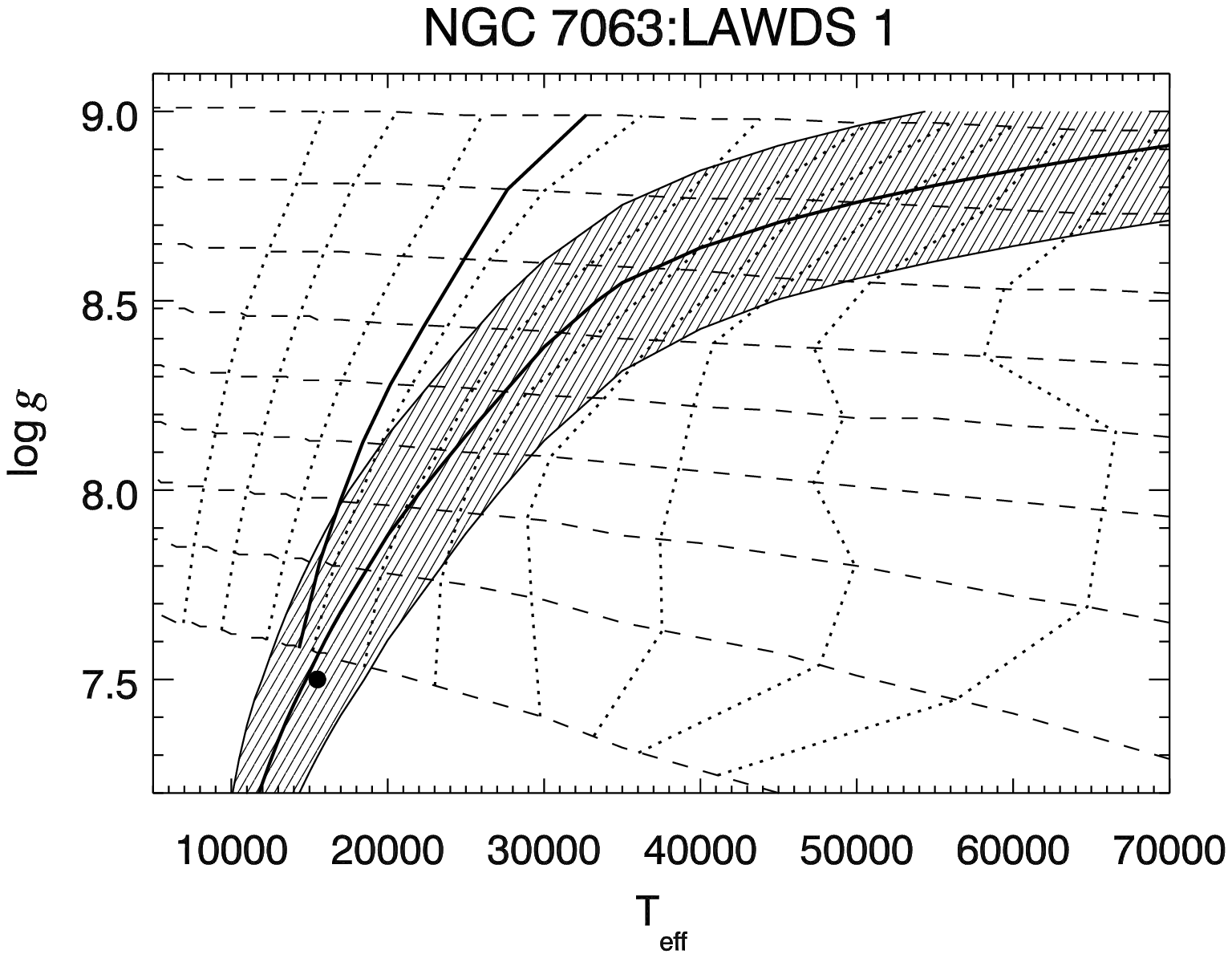}
\includegraphics[scale=0.45]{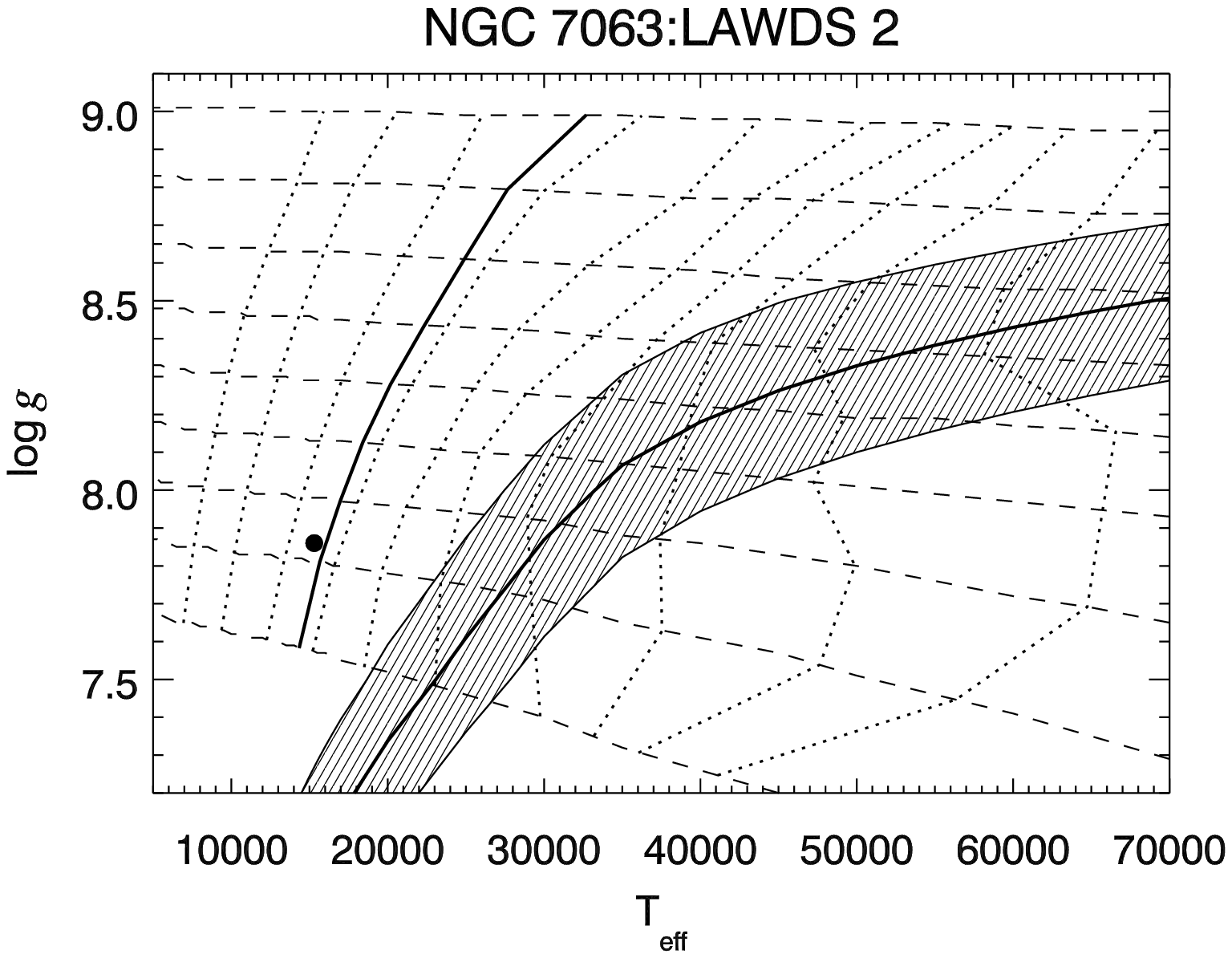}
\end{minipage}
\begin{minipage}{3.25in}
\includegraphics[scale=0.45]{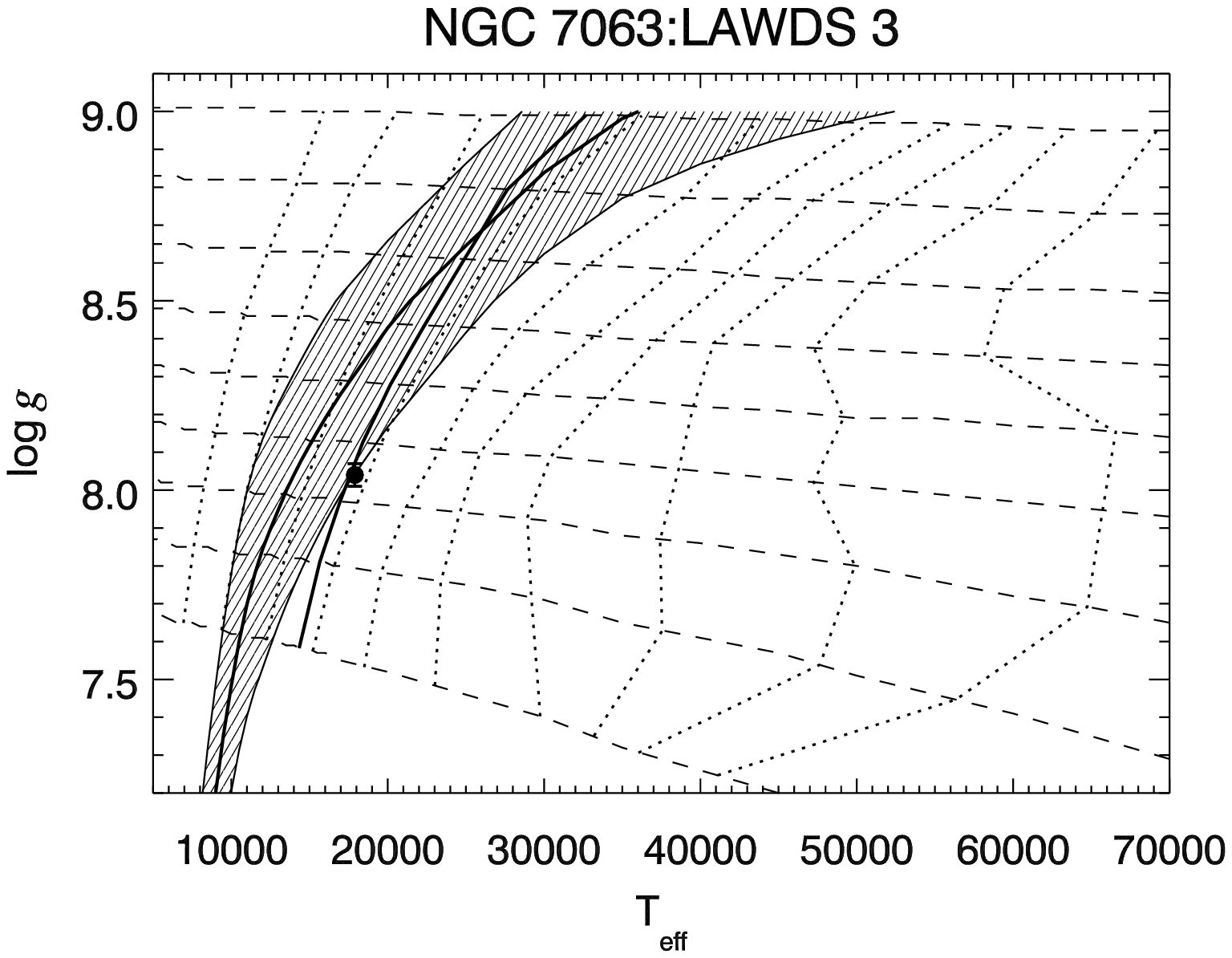}
\includegraphics[scale=0.45]{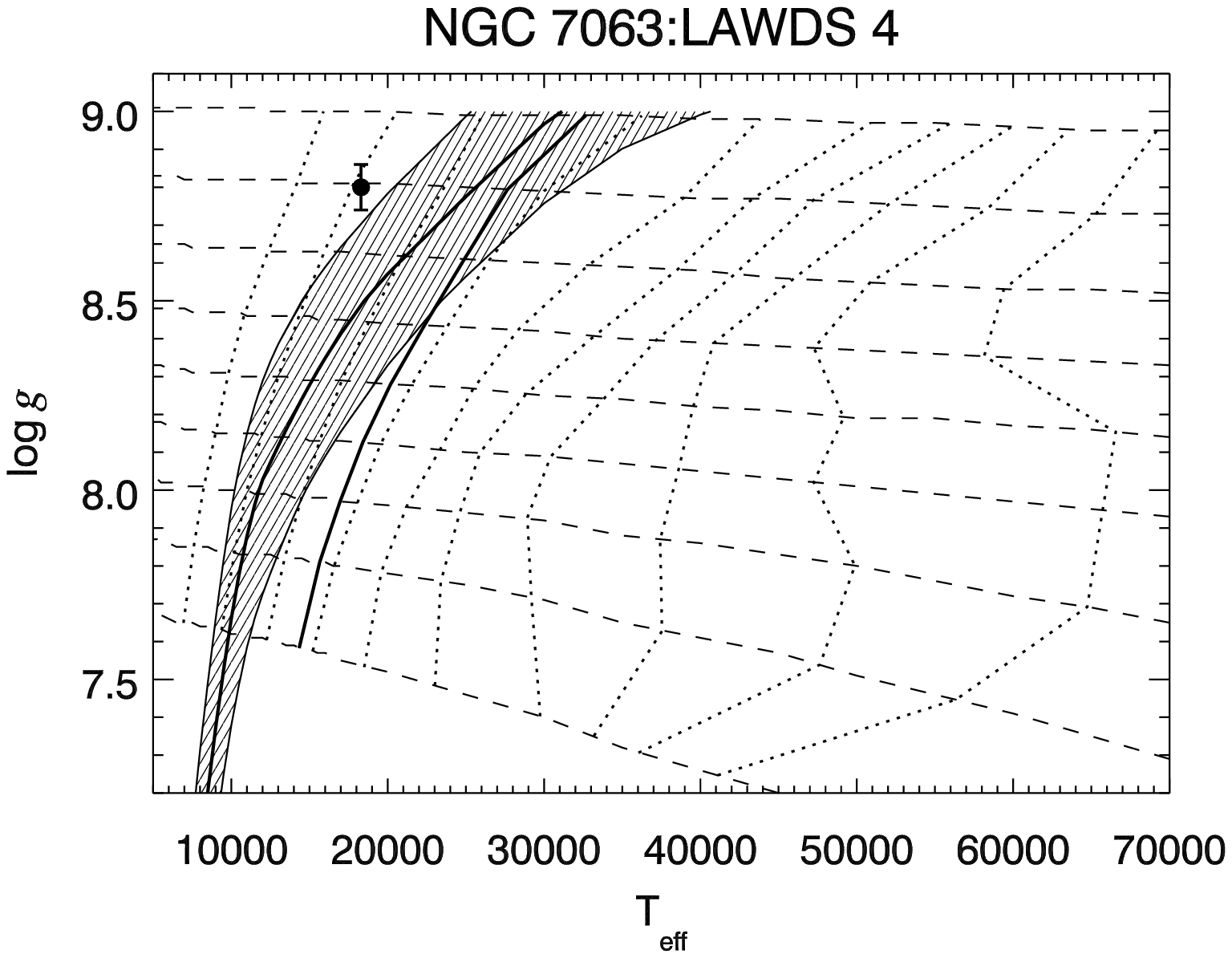}
\end{minipage}
\caption{Same as Figure \ref{fig.photdist_n6633}, except for WDs in
  the cluster NGC 7063.\label{fig.phot_dist.n7063}}
\end{figure*}

Cluster membership is determined based on the apparent distance
modulus $(m-M)_V$ and cooling age of each WD.  Potential cluster
members are defined as those with \tcool ~less than 1$\sigma$ above
the adopted cluster age and distance moduli within 2$\sigma$ of the
cluster distance.  For NGC 6633, these criteria are $\tcool\leq 625$
Myr and $8.34 \leq (m-M)_V \leq 8.70$. In the literature, no errors
are explicitly stated on the distance to NGC 7063, so we adopt the
scatter in published values of $\sim 0.2\,{\rm mag}$; the
corresponding selection criteria for NGC 7063 WDs are $\tcool\leq 158$
Myr and $ 9.07 \leq (m-M)_V \leq 9.87$.  Applying these criteria, we
find one cluster member in NGC 6633 (NGC 6633:LAWDS 27), though up to
three more members may exist; see \S\ref{sec.binarywds} and
\S\ref{sec.n6633_wb16}.  One cluster member in NGC 7063 (NGC
7063:LAWDS 1) is detected. A graphical exposition of this process is
shown in Figures \ref{fig.photdist_n6633} and
\ref{fig.phot_dist.n7063}.

We calculate progenitor masses for cluster member WDs using stellar
evolutionary tracks (including convective overshoot) from
\citet{2000A&AS..141..371G} and \citet{1994A&AS..106..275B}.  The
calculated \tcool ~is subtracted from the cluster age to produce the
progenitor star's lifetime from the zero-age main sequence (ZAMS)
through the planetary nebula phase.  The ZAMS mass corresponding to
the progenitor lifetime is then determined from the evolutionary
tracks.  As the cluster metallicities differ from those of the
evolutionary tracks, we linearly interpolate the progenitor masses to
the assumed metallicity.  Progenitor masses for the cluster WDs are
given in Table \ref{tab.minit}.

There may be some field WDs that meet our selection criteria for
cluster members. We estimate the expected number of interlopers as
follows.  The bright WD luminosity function from the Palomar-Green
sample of \citet{2005ApJS..156...47L} can be approximated by a power
law:
\begin{equation}
\log\phi = 0.84M_V - 13.3\,,
\end{equation}
where $\phi$ is in units of pc$^{-3}$ 0.5 mag$^{-1}$.  We integrate
this to limiting magnitudes of $M_V=12.5$ for NGC 6633 and $M_V=11.75$
for NGC 7063, both of these values approximating the luminosity of a
1.2\msun ~WD with a cooling age equal to the cluster age.  The
resulting field WD densities are $1.6\times 10^{-3}$ pc$^{-3}$ for NGC
6633 and $3.9\times 10^{-4}$ pc$^{-3}$ for NGC 7063.  The volumes in
which these field white dwarfs could reside are truncated pyramids
with faces defined by the image boundaries and bases defined by the
distance selection criteria for each cluster. Assuming all extinction
is foreground to the selection volume, these volumes are 1060 pc$^3$
for NGC 6633 and 520 pc$^3$ for NGC 7063. So, we expect an average of
1.7 field WDs in the NGC 6633 and 0.2 field WDs in the NGC 7063 to
meet the cluster member selection criteria.

Assuming Poisson statistics, there is an $\approx 18\%$ probability
that the NGC 6633 region has no field WDs, while there is an $\approx
82\%$ probability that NGC 7063 has no field WDs.  Based on these
relatively ambiguous statistics, we cannot claim with any confidence
whether the candidate cluster WDs are truly cluster members or are
just part of the field WD population.  Proper motion measurements will
be necessary to confirm either scenario.

\subsection{Notes on Individual Objects}

\subsubsection{NGC 6633:LAWDS 4 and NGC 6633:LAWDS 7 \label{sec.binarywds}}
These WDs are both relatively massive (0.79\msun ~and 0.87\msun,
respectively), as one might expect for NGC 6633 WDs such as NGC
6633:LAWDS 27 (0.77\msun).  These two WDs also have nearly identical
distances [$(m-M)_V=7.75\pm 0.02$ and $(m-M)_V=7.78\pm 0.03$], placing
them foreground to the cluster by 0.77 and 0.74 mag, respectively.
These two WDs therefore make strong candidates for binary WDs.

From our data, it is not yet possible to tell if one or both of these
objects is actually a double degenerate.  \citet{1989ApJ...345L..91B}
find that the unresolved binaries can be fit acceptably by a
single-star spectrum, with the resulting fit \logg ~and \teff
~intermediate to that of the two binary components.  The color of NGC
6633:LAWDS 4 ($\bv = 0.16\pm 0.03$) is what would be expected based on
the spectral parameters and the (reddened) photometric models of
\citet{Holberg06}: $(\bv)_{\rm model} = 0.11$, but the color of NGC
6633: LAWDS 6 ($\bv = 0.38\pm 0.03$) is redder than the models would
imply [$(\bv)_{\rm model} = 0.17$].  This redder color could suggest a
cooler component to the system.

The presence of binary WDs in open clusters has been predicted based
on dynamical models \citep[e.g.,][]{2003ApJ...589..179H}.  Further,
the presence of such binaries may be useful for explaining the
perceived deficit of WDs in some open clusters and for determining the
distribution of binary mass ratios \citep{2004ApJ...601.1067W}.

However, if these WDs are double-degenerate cluster members, the
binary components would have to have virtually identical luminosities
to appear 0.75 mag over-luminous, further implying that the two WDs
are likely of similar masses. Since WDs of different masses cool at
different rates, it would be a surprising coincidence if two separate
systems both happened to have different-mass components presently with
the same luminosity.

Further, if these are nearly equal-mass component binaries,
\emph{both} pairs would have a combined mass above the Chandrasekhar
mass, making these candidates for Type Ia supernova progenitors,
depending on their orbital separations.  However, the ESO Supernova Ia
Progenitor SurveY (SPY) has only detected 3 binaries with combined
masses over the Chandrasekhar limit out of over 100 detected binary
WDs \citep{2005ASPC..334..375N}.  We therefore emphasize that the
identification of these objects as double degenerates is, at present,
speculative.

It may be possible to determine whether these two objects are binaries
with additional, high-resolution spectroscopy of the H$\alpha$ line,
where two non-LTE line cores or a single core with variable velocity
should be detectable.  In addition, second-epoch deep imaging of this
cluster will allow proper motions to be measured and the probability
of cluster membership to be determined.

\tabletypesize{\footnotesize}
\begin{deluxetable*}{lccccc}
\tablecolumns{6}
\tablewidth{0pt}
\tablecaption{Cluster White Dwarf Initial Masses\label{tab.minit}}
\tablehead{\colhead{White Dwarf} & \colhead{$M_f$\tablenotemark{a}} & \colhead{$\sigma_{M_f}$\tablenotemark{a}} 
  & \colhead{$M_i$} & \colhead{Random Error} & \colhead{Systematic Error\tablenotemark{b}} \\
  & \colhead{(\msun)} & \colhead{(\msun)} & \colhead{(\msun)} & \colhead{(\msun)} & \colhead{(\msun)}}
\startdata
NGC 6633:LAWDS  4 & 0.79 &  0.01 & 2.93 &  0.01 & $^{+0.16}_{-0.14}$ \\
NGC 6633:LAWDS  7 & 0.87 &  0.01 & 3.20 &  0.02 & $^{+0.23}_{-0.20}$ \\
NGC 6633:LAWDS 27 & 0.77 &  0.01 & 3.30 &  0.03 & $^{+0.27}_{-0.22}$ \\
NGC 7063:LAWDS  1 & 0.37 &  0.01 & 8.17 &  0.16 & $^{+\infty}_{-1.99}$ \\
\enddata
\tablenotetext{a}{From Table \ref{tab.wdfits}}
\tablenotetext{b}{From uncertainty in cluster age}
\end{deluxetable*}

\subsubsection{NGC 6633:LAWDS 16\label{sec.n6633_wb16}}
The spectrum of NGC 6633:LAWDS 16 exhibits absorption features of
\ion{He}{1}, but shows no evidence for hydrogen.  We therefore
classify this object as a DB WD.  As of yet, we have not developed
code for fitting model atmospheres to DB spectra. We estimate the WD
temperature based on the equivalent width (EW) of He 4471\AA, as
calculated by \citet{1980A&AS...39..401K}. The EW of the He 4471\AA
~line (measured from 4350\AA ~to 4600\AA) is $\approx 18$\AA; this
translates to a temperature of $\sim 16000$ K for $\logg=8$.

It is also possible to use photometry to determine if this object is a
potential cluster member.  Assuming that NGC 6633:LAWDS 16 is at the
cluster distance and reddening, its photometry (see Table
\ref{tab.cands.n6633}) implies $M_V=11.56$ and $(B-V)_0=-0.025$.  In
Table \ref{tab.n6633wb16}, atmospheric parameters for DB WDs of
various masses with $M_V=11.56$ are calculated from data presented in
\citet{Holberg06}.  From the table, it is evident that the observed
photometry of NGC 6633:LAWDS 16 is consistent with a cluster DB WD of
mass $\sim 0.6\msun$ -- 0.8\msun, with \teff ~consistent with that
derived from the EW measurement.  In addition, Figure
\ref{fig.n6633_ids} shows the DB lying just above the 0.8\msun ~DB
cooling curve.

We therefore surmise that NGC 6633:LAWDS 16 could be a cluster member.
The likely progenitor mass of $\sim 3.5\msun$ (see Table
\ref{tab.n6633wb16}) is similar to that of the DA cluster member, and
if the WD mass is toward the upper end of the 0.6\msun--0.8\msun
~range, it would fall close to the empirical initial-final mass
relation.

If a cluster member, this would represent the third He-atmosphere WD
to be detected in an open star cluster, along with \object{LP 475-252}
and \object{NGC 2168:LAWDS 28} \citep{2006ApJ...643L.127W}.  As such,
this WD may be an important point in understanding the purported
difference in the ratio of DA to non-DA WDs in open clusters and in
the field.  Further observations of this object, both higher S/N
spectroscopy and proper motion determination, are needed to confirm
whether this DB is a cluster member.

We note that another detected WD in the field, NGC 6633:LAWDS 14, is
also a cool DB WD.  This object was shown to be foreground to the
cluster in \citet{1994A&A...285..451R}.

\subsubsection{NGC 7063:LAWDS 1}
The calculated distance modulus to NGC 7063:LAWDS 1 is
$(m-M)_V=9.61\pm 0.04$, less than 2$\sigma$ away from the cluster
distance modulus of $(m-M)_V=9.41$, and is therefore identified as a
cluster member.  However, its mass (0.37\msun) is very low for its
initial mass ($8.2^{+\infty}_{-2.0}\msun$).  We therefore conclude
that, if a cluster member, NGC 7063:LAWDS 1 is likely a He-core WD and
the product of binary evolution \citep[e.g.,][]{1993PASP..105.1373I}.
If this hypothesis is correct, then follow-up observations may be able
to detect the companion \citep[e.g.,][]{1995MNRAS.275..828M} or
evidence for an unseen companion, such as radial velocity variations
in the core of H$\alpha$.  This system could then be useful for
constraining various time scales in common-envelope evolution.

\subsubsection{NGC 7063:LAWDS 6\label{sec.n7063wb6}}

Our spectral fits to NGC 7063: LAWDS 6 give $\teff=11,100$K and
$\logg=7.00$, results typical of attempting to fit A stars with our
routine.  However, qualitative analysis of the spectrum suggests that
the surface gravity is higher than $\logg=7$.  This discrepancy is due
to a nearby bright star in the slit hampering the sky subtraction to
such a degree that a reliable fit cannot be obtained.

 If NGC 7063:LAWDS 6 is a cluster member, it would have $M_V=12.5$ and
$(B-V)_0=0.33$.  The range of effective temperatures for a WD of this
$M_V$ goes from 8000K (for $\mwd=0.4\msun$) to 13,000K (for
$\mwd=1.0\msun$), with corresponding cooling ages of 700 Myr and 834
Myr, respectively. If the WD is very high mass (1.2\msun), then this
$M_V$ corresponds to $\teff=20,000$K and $\tcool=511$Myr.  The color
favors lower \teff ($\sim 7250$K).  It therefore appears likely that
this WD is older than the star cluster and so not a cluster member.

\subsection{Expected number of cluster white dwarfs}
The number of expected cluster WDs can be estimated by normalizing an
assumed IMF to the number of known cluster stars.  Based on the proper
motion study of NGC 6633 by \citet{1973A&AS....9..213S}, the cluster
has 45 likely members ($P\geq 50\%$) brighter than $V=12$,
corresponding to $M_V\approx 3.5$.  By assuming a Salpeter IMF, an
upper mass limit to WD progenitors of $8\msun$, and by using the Monte
Carlo calculations described in \citet{2004ApJ...601.1067W}, we
calculate that $7.8\pm 3.0$ WDs should be detectable in NGC 6633, one
of which should be binary.  This compares with $\leq 4$ cluster WDs
found in this paper. The same calculation for NGC 7063, which has 17
likely member stars brighter than $M_V\approx 4.7$ in the portion of
the field imaged by PFCam, results in $1.1\pm 1.1$ cluster WD being
detectable in the field.

Thus, like the Hyades, NGC 6633 shows a likely deficit of WDs that
cannot be explained by WDs being hidden in binaries.  The NGC 6633
imaging contains almost the entire cluster; the CFH12K field covers
$42\arcmin\times 28\arcmin$ centered on the cluster core, and
\citet{1973A&AS....9..213S} detects proper motion cluster members over
a diameter of $\approx 40\arcmin$.  However, it is possible that some
number of WDs lie outside our imaged area.  This areal coverage issue
has been found to be the likely cause of the apparent deficit of WDs
in Praesepe \citep{2004MNRAS.355L..39D,2006MNRAS.369..383D}.

\subsection{The initial-final mass relation}
\begin{figure}
\plotone{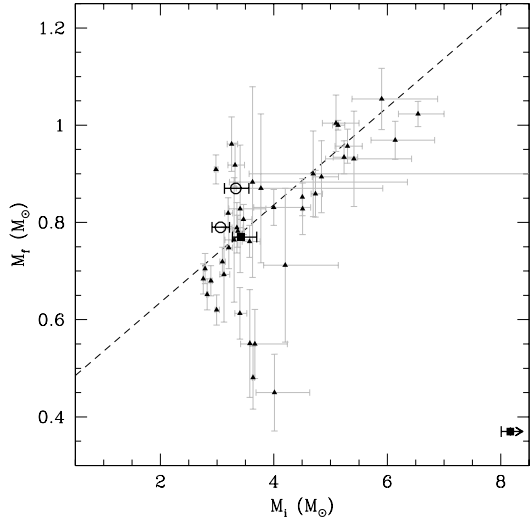}
\caption{The empirical initial-final mass relation.  Filled triangles
  are from \citet{2005MNRAS.361.1131F}; the gray error bars do
  \emph{not} include the uncertainties in cluster ages. The dashed
  line is the linear fit to the IFMR from that work.  The filled
  squares are the two cluster DAs from this work; the open circles are
  the potential double-degenerate binaries.  Errors in final mass are
  smaller than the point size and are therefore not shown.
  The indicated errors in initial mass are 1$\sigma$ ~errors
  in the cluster ages.\label{fig.ifmr}}
\end{figure}

If we assume that both candidate cluster DA WDs are indeed members of
their respective clusters, we can place these two objects on the
empirical IFMR.  In Figure \ref{fig.ifmr}, we plot these two points
along with the data collected by \citet{2005MNRAS.361.1131F}.  NGC
6633:LAWDS 27 is seen to fall on the \citet{2005MNRAS.361.1131F}
relation, while NGC 7063:LAWDS 1 falls well off the relation.  This
can be readily explained if NGC 7063:LAWDS 1 is the product of binary
evolution (see \S\ref{sec.n7063wb6}) or not a true cluster member.
NGC 6633 has a lower metallicity than the Hyades and Praesepe, yet the
NGC 6633 WD occupies the same region of the initial-final mass
relation as the WDs in the Hyades and Praesepe.  This suggests that
metallicity, at least over this small metallicity range, has little
impact on the WD mass, though the observational scatter is still quite
large and can easily swamp a small signal.

\subsection{Three DAZ White Dwarfs}
At least three WDs in this sample show potential \ion{Ca}{2} K
absorption: NGC 6633:LAWDS 8 (EW$=0.6$ \AA), NGC 6633:LAWDS 15
(EW$=0.4$\AA), and NGC 6633:LAWDS 27 (EW=$0.6$\AA).  While these
features could be interstellar in nature, the other WDs in the cluster
field at similar or greater distances have no discernible \ion{Ca}{2}
absorption. \citet{2003ApJ...596..477Z} determine that $\sim 25\%$ of
DA WDs show \ion{Ca}{2} K absorption, if the spectra are of high
enough S/N and sufficiently high resolution.  Our detection of
\ion{Ca}{2} in 3 of 11 DA WDs is fully consistent with that number.
We therefore classify these three objects as spectral type DAZ.

The origin of the metals in DAZ WDs remains controversial, with
explanations including accretion from the interstellar medium
\citep{1979A&A....72..104W}, cometary impacts
\citep{1986ApJ...302..462A}, and accretion of asteroidal material
\citep{1990ApJ...357..216G}.  However, recent data from the Spitzer
Space Telescope and the IRTF reveal that several DAZs with high $[{\rm
Ca/H}]$ have circumstellar debris disks \citep{Reach2005,Kilic2006}.
Two of these DAZs, NGC 6633:LAWDS 8 and NGC 6633:LAWDS 27, have \teff
~in the range where a debris disk may be detected.

\section{Conclusions}
We have obtained photometric and spectroscopic data for white dwarfs
(WDs) in two sparse open clusters, NGC 6633 and NGC 7063.  For the
open cluster NGC 6633, we determine a main-sequence turnoff age of
$560^{+70}_{-60}$ Myr.  We select 32 candidate WDs based on \bv~
photometry.  Spectroscopic follow-up of 22 of these candidates finds 9
of these to be WDs, 7 with spectral types DA or DAZ and 2 of spectral
type DB.  One DA and one DB have distance moduli consistent with
cluster membership.  Two other DAs are $\approx 0.75$ mag over-luminous
for cluster member WDs; these are potential double-degenerate cluster
WDs which, if truly binary, would each have combined masses above the
Chandrasekhar mass.  The NGC 6633 DB WD is the third known cluster WD
with a He-dominated atmosphere.

For the open cluster NGC 7063, we estimate a main-sequence age of
$125^{+33}_{-25}$ Myr, assuming solar metallicity.  Using \ubv
photometry, we identify nine candidate WDs.  Spectroscopy of 7
candidates confirms 5 WDs, though only one is consistent with cluster
membership.  This WD has a mass of 0.37\msun, and so is likely a
He-core WD resulting from binary evolution of two massive stars, with
the WD's progenitor mass $M = 8.17^{+\infty}_{-2.0}\msun$.

\tabletypesize{\scriptsize}
\begin{deluxetable}{ccccccc}
\tablecolumns{7}
\tablewidth{0pt}
\tablecaption{Photometrically-derived Quantities for NGC 6633:LAWDS 16\label{tab.n6633wb16}}
\tablehead{\colhead{Assumed \mwd} & \colhead{$(\bv)_0$} & \colhead{\teff} 
  & \colhead{\logg} & \tcool & \colhead{$M_i$} & \colhead{$\sigma_{M_i}$} \\
  \colhead{(\msun)} & & \colhead{(K)} & & \colhead{(Myr)} & \colhead{(\msun)} 
  & \colhead{(\msun)}}
\startdata
0.4 & $\phantom{-}0.059$ & 11000 & 7.660 & 322 &   3.72 & $^{+0.47}_{-0.34} $\\
0.5 & $\phantom{-}0.008$ & 12215 & 7.868 & 305 &   3.62 & $^{+0.41}_{-0.31} $\\
0.6 & $-0.026$ & 13341 & 8.017 & 310 & 3.64 & $^{+0.43}_{-0.32} $\\
0.7 & $-0.056$ & 14656 & 8.163 & 286 & 3.52 & $^{+0.36}_{-0.28} $\\
0.8 & $-0.076$ & 16293 & 8.320 & 271 & 3.45 & $^{+0.33}_{-0.26} $\\
0.9 & $-0.088$ & 18469 & 8.480 & 250 & 3.36 & $^{+0.29}_{-0.24} $\\
1.0 & $-0.110$ & 21973 & 8.650 & 205 & 3.20 & $^{+0.23}_{-0.20} $\\
1.1 & $-0.158$ & 30801 & 8.830 & 102 & 2.92 & $^{+0.15}_{-0.14} $\\
\enddata
\end{deluxetable}

The NGC 6633 DA WD and both potential double degenerates are
consistent with the existing empirical initial-final mass
relation. Most of the points in the initial-final mass relation in the
region occupied by the NGC 6633 WD(s) are from the super-solar
metallicity open clusters Praesepe and the Hyades, so the agreement of
the sub-solar metallicity NGC 6633 WD(s) with these points suggests
little metallicity dependence in the initial-final mass relation over
this metallicity range.

Three of our 11 DA WDs show \ion{Ca}{2} absorption lines and are
therefore classified as spectral type DAZ.  The ratio of DAZs to DAs
is consistent with other published studies indicating that $\sim 25\%$
of hydrogen-atmosphere WDs are DAZs.

One open issue is the membership of the reported cluster WDs.  While
the observed distance modulus is an important diagnostic, we cannot
rule out that the cluster WDs are interlopers from the field.  An
independent method of determining cluster membership, such as proper
motion measurements for these WDs, would greatly assist in this
regard.

\acknowledgements \emph{Acknowledgements} --- 
The authors are grateful for financial support from
National Science Foundation grant AST 03-07492.  KAW is supported by
an NSF Astronomy and Astrophysics Postdoctoral Fellowship under award
AST-0602288. We thank the anonymous referee for thorough comments that
helped us to improve this paper.  We would also like to thank James
Liebert for many extremely helpful discussions during the preparation
of this paper and for attempting to share his vast accumulation of
white dwarf wisdom with us.  We also thank Liebert, along with Pierre
Bergeron and Jay Holberg, for sharing their PG white dwarf data with
us; we warmly thank Detlev Koester for providing us with his white
dwarf atmospheric models; and we thank Matt Wood and Gil Fontaine for
providing their white dwarf evolutionary models (obtained from the
latter through Bergeron).  We heartily thank Jason Harris for
providing us with pieces of his StarFISH code.
Some data in this paper were obtained as guest users of the Canadian
Astronomy Data Centre, which is operated by the Dominion Astrophysical
Observatory for the National Research Council of Canada's Herzberg
Institute of Astrophysics.  This research has made use of the WEBDA
database, operated at the Institute for Astronomy of the University of
Vienna.
The authors wish to recognize and acknowledge the very significant
cultural role and reverence that the summit of Mauna Kea has always
had within the indigenous Hawaiian community.  We are most fortunate
to have the opportunity to conduct observations from this mountain.

\end{document}